\documentclass[tradiabstract]{aa} % for the abstract without structuration
\usepackage{graphicx}
\usepackage{longtable}
\usepackage{natbib}
\bibpunct{(}{)}{;}{a}{}{,} % to follow the A&A style
\usepackage{lscape}
\usepackage{txfonts}
\begin{document}

   \title{Impact of
  nonconvergence and various approximations of
  the partition function on the molecular column densities in the
  interstellar medium~\footnote{Tables D.1, D.2 and D.3 are only available in electronic form
at the CDS via anonymous ftp to cdsarc.u-strasbg.fr (130.79.128.5)
or via http://cdsweb.u-strasbg.fr/cgi-bin/qcat?J/A+A/}}

%   \subtitle{I. Overviewing the $\kappa$-mechanism}

   \author{M. Carvajal \inst{1} \and C. Favre \inst{2,3} \and I. Kleiner \inst{4} 
\and C. Ceccarelli \inst{3} \and E. A. Bergin \inst{5} \and D. Fedele \inst{2}%\fnmsep\thanks{Just to show the usage
%          of the elements in the author field}
          }

   \institute{Dept. Ciencias Integradas, Facultad de Ciencias
  Experimentales, Centro de Estudios Avanzados en F\'{\i}sica,
  Matem\'atica y Computaci\'on, Unidad Asociada GIFMAN, CSIC-UHU,
  Universidad de Huelva, Spain\\
  Instituto Universitario ``Carlos I'' de F\'{\i}sica
  Te\'orica y Computacional, Universidad de Granada, Granada, Spain\\
              \email{miguel.carvajal@dfa.uhu.es}
              \and
              INAF--Osservatorio Astrofisico di Arcetri, Largo
              E. Fermi 5, Firenze, 50125, Italy
              \and
         Univ. Grenoble Alpes, CNRS, IPAG, F-38000 Grenoble, France
         \and
         Laboratoire Interuniversitaire des Syst\`emes Atmosph\'eriques
         (LISA) , UMR CNRS 7583,  Universit\'e Paris-Est Cr\'eteil,
         Universit\'e de Paris, Institut Pierre Simon Laplace (IPSL) ,
         Cr\'eteil, France 
         \and
         Dept. of Astronomy, University of Michigan, 311 West Hall, 1085 South University Avenue, Ann Arbor, MI 48109, USA
   }

   \date{Received ..., 2019; accepted ... , 2019}

% \abstract{}{}{}{}{}
% 5 {} token are mandatory

  \abstract
  % context heading (optional)
  % {} leave it empty if necessary
   {We emphasize that the completeness of the partition function,
that is, the use of a converged partition function at the typical
  temperature range of the survey, is very important to decrease the
  uncertainty on this quantity and thus to derive reliable
  interstellar molecular densities.  
In that
context, we show how the use of different approximations for the
rovibrational partition function together with some interpolation
and/or extrapolation procedures may affect the estimate of the
interstellar molecular column density. For that purpose,
we apply the partition function calculations to astronomical
observations performed with the IRAM-30m telescope towards the
NGC7538--IRS1 source of two N-bearing molecules:
isocyanic acid (HNCO, a quasilinear molecule) and methyl cyanide
(CH$_3$CN, a symmetric top molecule). The case of methyl formate
(HCOOCH$_3$), which is an asymmetric top O-bearing molecule
  containing an internal rotor is also discussed. Our analysis shows that the use of different partition
function approximations leads to relative differences in the
resulting column 
densities in the range 9 to 43\%. 
Thus, we expect this work to be relevant for surveys of sources with temperatures higher than 300~K  and to observations in the infrared.
}


\titlerunning{Partition function for the ISM surveys}

\authorrunning{M. Carvajal \inst{1} \and C. Favre \inst{2,3} \and I. Kleiner \inst{4} \and C. Ceccarelli \inst{5} \and E. A. Bergin \inst{6} \and D. Fedele \inst{2}}

   \maketitle
%
%________________________________________________________________

\section{Introduction}

Molecular astronomy needs accurate spectral analysis of the emission
associated with various molecular 
species in order to identify them and to estimate the physical
conditions of the interstellar region they are emitting from 
~\citep{herbst2009}. A comprehensive molecular spectral
characterisation is therefore extremely important for astrochemistry because the
 relative isotopic abundance
estimates, the branching ratios, and the rate coefficients along with
the activation energy of chemical reactions strongly relies on it
\citep[see][]{shaw}. In particular, a precise 
estimation of the column density of the molecular interstellar species
is required in order to investigate the possible chemical
reactions taking place in the interstellar medium (ISM).  
Such estimates are based on molecular spectroscopy that provides transition frequencies,
line strengths, and partition functions for a
given molecule 
  through accurate laboratory spectral analyses. 

Furthermore, the determination of ISM molecular and isotopic
abundance ratios provides strong insight into the 
molecular formation mechanisms that occur in the ISM. {To achieve it, it
  is necessary that}
the intensity calculation of the molecular species at different
temperatures is reliable.
It is important to note that the spectroscopic
determination of 
the transition frequencies, the line strengths, and the
partition function 
has to be very accurate due to the
high spectral resolution that is now accessible with the present
astronomical observatories.

 The advent of new infrared and (sub-)millimeter observatories
 in the last decade { (e.g., Atacama Large Millimeter Array (ALMA),
   Herschel, Stratospheric Observatory For Infrared Astronomy (SOFIA))} 
has motivated the molecular spectroscopy community to characterize
increasingly complex
molecules  for which spectra were unrecorded until then. 
The spectroscopic data are gathered
 through intensive laboratory work, both
 experimental and theoretical, to predict new and accurate
 molecular data in the spectral range covered by the observational
 instruments. In addition, these data permit the exploration of new frequency
   ranges and enable the  prediction of the frequencies via theoretical modeling.

Present databases 
compile and maintain the spectroscopic data updated in
catalogs commonly used by
the molecular astronomy community, such as the Cologne
Database for Molecular Spectroscopy 
(CDMS)\footnote{http://www.ph1.uni-koeln.de/vorhersagen/}~\citep{endres2016},
 the JPL (Jet propulsion Laboratory) database\footnote{https://spec.jpl.nasa.gov/}~\citep{pickett1998}, the Lovas/NIST
catalog\footnote{http://physics.nist.gov/restfreq}~\citep{lovas2004}, the Toyama Microwave Atlas for spectroscopists and
astronomers\footnote{http://www.sci.u-toyama.ac.jp/phys/4ken/atlas/},
the SPLATALOGUE database\footnote{This database also collects
  data from other catalogues,
  http://www.splatalogue.net/}~\citep{remijan2007} and
  HITRAN\footnote{https://hitran.org/}~\citep{gordon2017,gamache2017}. These databases
have compiled a huge 
amount of data provided by spectral analyses performed via intensive laboratory
spectral recordings. 

So far, spectroscopic studies have made possible the 
identification of about 200 molecular
species\footnote{http://www.astro.uni-koeln.de/cdms/molecules/} in star-forming regions and in the
ISM. 
Nevertheless, for a number of molecular species, some reported physical quantities are not
  always normalized among the different authors. 
 This is the case for example for the partition function values:  
  there are several definitions of the nuclear spin
  statistical weight and  the partition function is not always accounted for. This is also the case of 
  the line strengths: some authors use a definition involving the square
  molecular dipole moment while others do not.

Moreover, the internal partition functions
can be computed in different ways: A direct sum formula can be used,
which involves the exponential of the energy levels, if those energy
levels are known; if they are not known one can use various
approximations to get the partition function. {The main issue} is the
uncertainty on the partition function and its effect on molecular
column densities. When the partition function is computed with the
direct sum formula, sometimes it is provided without carrying out an
appropriate convergence study in the temperature ranges of the ISM,
typically from 9.375  to 300 K. The convergence on the partition
function is said to be reached when a complete (full) list of
rovibrational energy levels is available at the temperature of a given
survey. In that case, due to the integrative nature of partition
functions, completeness is more important in general than the accuracy
with which those energy levels are
estimated~\citep{furtenbacher2016}. On the contrary, when 
the partition function is computed using various levels of
approximations (because the energy level information is not or not
easily available), large uncertainties on the partition function can
also occur.  

Molecular column densities (number of molecules per unit area along
the line of sight, cm$^{-2}$) are traditionally retrieved by a population
rotational diagram \citep{mangum2015}. Since the total molecular
column density is proportional to the internal partition function, the
uncertainties on the partition function will directly affect it.  In
particular, using an uncomplete (thus smaller) partition function will
lead to underestimation of the molecular column densities and therefore
some astrochemical conclusions could turn out to be slightly or
even significantly different. 

\citet{mangum2015} published a review aimed at 
describing how to calculate the molecular column density from
  molecular spectral (rotational or ro-vibrational)
  transitions. Some years before, \citet{fischer2002}
  and \citet{fischer2003} studied, for atmospherical and astrophysical
  species, the convergence of the internal
  partition function. However none of these former studies focused on
  the implications of using an approximate (or a not fully converged)
  partition function on the estimates of the interstellar molecular
  column densities.

The present paper aims at addressing the impact of different levels of
approximations of the partition function on estimations of the interstellar molecular
physical conditions. 
Some partition function interpolation and extrapolation
 procedures commonly used in the  
literature are also presented 
along with the analysis of their relevance regarding the
  temperature range. The interpolation and extrapolation procedures
    are indeed often used to determine the partition function value
  at any given ISM temperature.
The effect of using an incomplete, that is, not fully converged, partition function is illustrated below via the use of 
the three following molecules, which represent different molecular
geometries: isocyanic acid (HNCO), a quasilinear molecule;  methyl
cyanide (CH$_3$CN), a symmetric top molecule;  and methyl formate
(HCOOCH$_3$), an asymmetric top molecule with a large amplitude
internal rotor. Finally, a fourth molecule, hydrogen sulfide
  (H$_2$S), serves us to illustrate the effect of 
  anharmonicity on the vibrational contribution of the partition
  function.

This paper is organized as follows: in Section~\ref{sec-rotdiag} 
we briefly recall how to estimate the temperature and the 
interstellar molecular abundance from the molecular spectra under the
assumption of local thermodynamic equilibrium (LTE). 
In Section~\ref{sec-rotdiag} we highlight the need for a
complete, that is to say a convergent, partition function to provide more 
accurate {estimates and} to decrease the uncertainties of the molecular column densities.
In Section~\ref{sec-Q}, 
we outline the various approximations that can be done 
for the rovibrational partition function calculation and 
the various interpolation and extrapolation procedures in terms of temperature. Section~\ref{sec-astro}
describes the 30 m astronomical observations of NGC7538-IRS1. 
In Section~\ref{sec-discussion}, we  
give examples of derived molecular column
density estimates from different partition function
approximations or interpolation and extrapolation procedures.
Finally, the conclusions are set out in Section~\ref{sec-conclusions}.

\section{Relevance of a suitable partition function convergence study in the
  rotational temperature diagram estimates}
\label{sec-rotdiag}
In this section, we briefly describe how the ISM
molecular column densities and/or abundances are derived. 
For emission lines associated with a given molecule in an astronomical
survey, one can derive the total molecular column density $N$ as a
function of the integrated intensity $W$, the partition function and
the excitation temperature. The mathematical expression depends on the
assumption considered \citep[i.e., optically thin emission, negligible
background temperature, LTE, and so on; for
further details see][]{goldsmith1999,mangum2015}.

We therefore assume that LTE is reached and the
excitation temperature will be given by the quantity $T$. Hence, 
  in absence of collision rates, the
upper energy state column density, $N_{\rm u}$, can be related to the total
column density, $N$, of the molecule by  

  \begin{equation}
    \label{Nu}
N_{\rm u} = g_{\rm u} \frac{N}{Q(T)} \, e^{-\frac{E_{\rm u}}{k T}} 
  ,\end{equation}  

  \noindent where  $k$ is the Boltzmann constant, $Q(T)$ is the molecular
  partition function, and $g_{\rm u}$ and $E_{\rm u}$ are the degeneracy and the energy
  for the upper level u involved in the molecular transition,
   respectively. The degeneracy for the upper level involved in
    the transition $g_{\rm u}$ can be given as

  \begin{equation}
    \label{gu}
g_{\rm u}=(2J_{\rm u}+1) ~ g_{\rm ns}^{({\rm u})} ,
\end{equation}

\noindent where $J_{\rm u}$ is the upper state rotational angular momentum
and $g_{\rm ns}^{({\rm u})}$ is the nuclear spin statistical weight of the upper
state.

The total molecular column density $N$ of a molecular species, which
is related to the upper energy state column density $N_{\rm u}$ by Eq.~(\ref{Nu}), and
  the rotational temperature $T$ can be given by~\citep{goldsmith1999}:  

\begin{equation}
\label{logInt}
\ln \left[\frac{3 \, k \, W}{8 \, \pi^3 \, \nu \, S\mu^2 \, g_{\rm u}}
\right]= \ln \left[\frac{N}{Q(T)}\right] -
  \frac{E_{\rm u} }{k \, T}  ~~, 
\end{equation}

\begin{equation}
\label{logInt2}
\ln \left[\frac{8 \, \pi \, k \, \nu^2 \, W}{g_{\rm u} \, h \,
      c^3 \, A_{\rm ul}}
\right]= \ln \left[\frac{N}{Q(T)}\right] -
  \frac{E_{\rm u} }{k \, T}  ~~, 
\end{equation}

\noindent where $h$ is the Planck constant and $c$ is the speed
of light in vacuum. In the above  equations the integrated
  intensity $W$, given in {K$\cdot$km/s,} and transition frequency $\nu$ for a given line are
  provided through the spectral interstellar observations, and therefore
  their accuracies are limited by observational
constraints. The line strength $S\mu^2$ or the Einstein
coefficient A$_{\rm ul}$ for spontaneous emission, the
upper state energy $E_{\rm u}$  and  the molecular partition
function $Q(T)$, are all predicted after the spectral characterization of
the molecule is performed. Therefore, their accuracies are
also limited by the calculation
and prediction error bars based on laboratory work. 

To determine the column density $N$ and the temperature $T$ from the observations, a linear
regression, so-called {\em rotational temperature diagram}, is
carried out after substituting in Eq.~(\ref{logInt}) \textit{(i)}  the 
data for the detected lines from the observational survey namely $W$ and $\nu$, and \textit{(ii)} the spectroscopic data of
the molecules: $S\mu^2$, $g_{\rm u}$, $E_{\rm u}$, and $Q(T)$. 

The linear
regression of Eq.~(\ref{logInt}) provides the  y-intercept, which
corresponds to $b=\ln(N/Q)$, and the slope which is related to
the inverse of temperature by $a=-1/T$~\citep{turner1991}. Therefore, the
  excitation temperature can easily be
  estimated from the slope whereas
  the total molecular column density $N$ is obtained from the
  y-intercept 
provided that the molecular partition function is
  known. This means that the molecular partition function values at a
  characteristic temperature range directly affect the
  value of $N$. This is why a detailed convergence study of the
molecular partition function is relevant to obtain a
  correct estimate of $N$. 

For instance, if the source under study is a hot
  region where excited vibrational states of a given
    molecule are populated ($T\ge$300~K) and if the molecular
  partition function is computed only taking into account the rotational structure
  of the molecule (i.e., in its vibrational ground state), the resulting molecular column
  density will be underestimated. Therefore, for high-temperature
  regions, the rotational temperature diagrams should contain
    both the rotational and vibrational contributions and, therefore
    a more accurate designation for them should be rovibrational temperature diagrams rather than rotational
    temperature diagrams, which is a slightly misleading term in some cases.

In Section~\ref{sec-astro}, the rovibrational temperature
  diagrams for CH$_3$CN, HNCO, and HCOOCH$_3$ resulting from our
  observations are given as examples.

\section{Molecular partition function}
\label{sec-Q}

In this section, we present some approximations used for the
  calculation of the rovibrational partition function of a free
    molecule with the aim of
  reaching the correct values (under a convergence study) at the typical temperature range of the ISM.

A molecular partition function\footnote{Also referred to as {\em Total Internal
    Partition Sums} (TIPS)~\citep{fischer2002} and {\em ideal-gas
      internal partition function}~\citep{furtenbacher2016}.} is defined as a direct sum of exponential
terms that involve the energy levels and the temperature of the
environment where the molecule is located. Therefore, the direct
sum of the partition function can be given as

\begin{equation}
\label{rtvQ}
{Q(T)= \sum_i  g_{\rm ns}^{({\rm i})} ~(2 \, J_{\rm i} + 1) \, e^{-{E_{\rm i} \over k T}}  ,}
\end{equation}

\noindent where $J_{\rm i}$ and $g_{\rm ns}^{({\rm i})}$ are the rotational angular
momentum and the nuclear spin degeneracy of
    the energy level i, respectively, and $E_{\rm i}$ is the
    rovibrational energy 
    usually referred to the ground vibrational state as it is
assumed that uniquely the ground electronic state is
populated~\citep{fischer2002,cerezo2014}. This means that an
accurate partition function will be determined at any temperature
if all energies $E_{\rm i}$ are accurately known. 
At the present time, this is unfortunately often not possible
because all transitions
(and thus all energy levels) have not 
  yet been measured. From a quantum mechanical point of view, one might assume that the molecular
  parameters derived from the observed transition frequencies can be
  used to predict the missing energy levels, however
  their estimates are not always lying within the experimental
  uncertainty. Therefore, the exponential terms in Eq.~(\ref{rtvQ}) are 
limited to the lowest energies $E_{\rm i}$ up to a certain upper
threshold. 

The energy threshold is selected depending on the
environment under study, for example, ISM or planetary atmospheres,
which correspond to given temperature ranges. Regarding the number of
exponential terms in  Eq.~(\ref{rtvQ}), once the
  numerator $E_{\rm i}$ is large in comparison with the denominator $k T$,
  the exponential $e^{- {E_{\rm i} \over k T}}$ can be neglected
  as long as $E_{\rm i}$  reaches a certain upper threshold. This means that
according to the temperature of the environment, the upper threshold energy
can be selected so that the direct sum in  Eq.~(\ref{rtvQ}) uses
  a complete set of rovibrational energy levels and therefore reaches
convergence~\citep{fischer2002}.

In general, when the energy levels do not appear in { a database} or in the
literature, or cannot be calculated easily, approximations are usually
considered for determining the molecular partition 
function. In the case of the C$_2$ molecule, the internal partition
function is computed via the direct sum formula using a combination
of experimental and theoretically derived energy
levels~\citep{furtenbacher2016ApJS}. The theoretical treatment allows
the authors to obtain a full set of energies for the nine electronic
states considered during the determination of the ideal-gas
thermochemistry of C$_2$. However, those theoretically derived
rovibronic levels are complete but of limited accuracy. The sources of
inaccuracies and uncertainties in the partition function are fully
discussed in this paper. Unfortunately this is not always the case in
the literature. 
We investigate here the particular case of molecules that
possess one large 
amplitude torsional mode arising from a CH$_3$ top. In that instance,
$E_{\rm i}$ stands for the 
vibrational-torsional-rotational energy (also referred to as the
vibrational-torsional ground state) and the partition
  function can be approximated as a product of the rotational
contribution $Q_{\rm rot}(T)$, the torsional $Q_{\rm tor}(T),$ and the vibrational
$Q_{\rm vib}(T),$~\citep{Herzberg}, assuming that the torsional-vibrational-rotational coupling
interactions in the Hamiltonian can be left out:

 \begin{equation}
\label{rtvQapprox}
Q(T) \approx Q_{\rm rot}(T) \, Q_{\rm tor}(T) \, Q_{\rm vib}(T) ~~.
\end{equation}

\noindent In this case, for some
  isotopologs of methyl formate, \citet{favre2014} assessed that
  Eq.~(\ref{rtvQapprox}) is a good approximation for temperatures under 300~K.

If the
molecule has no torsional mode, the torsional contribution $Q_{\rm tor}$
is simply omitted in Eq.~(\ref{rtvQapprox}). 

\subsection{Rotational partition function}
The rotational partition
function $Q_{\rm rot}(T)$ (e.g., \citealt{Herzberg,groner2007}) can be written as
a direct sum,

\begin{equation}
\label{Qrot}
Q_{\rm rot}(T) = \sum_{\rm i}  g_{\rm ns}^{({\rm i})} \, (2 \, J_{\rm i}
+1) \, e^{-{E^{({\rm rot})}_{\rm i} \over k
    T} 
}
,\end{equation}

\noindent where $E^{({\rm rot})}_{\rm i}$ represents the energy for the i-th
rotational state. The degeneracy $g_{\rm ns}^{({\rm i})}$ is
included in the definition of the 
  rotational contribution of the partition function
  (Eq.~\ref{Qrot}) because in general $g_{\rm ns}^{({\rm i})}$
    is associated with the K degeneracy of the symmetric top (or
    $K_{\rm a}$
    and $K_{\rm c}$ for asymmetric top; see Sect.~\ref{subsec-gu}). In the
    present paper we address the special case of symmetric (or
    asymmetric) tops where
    the degeneracy  is state independent, meaning that  $g_{\rm
      ns}^{({\rm i})}=g_{\rm ns}$.

The classical approximation of the rotational partition function (\ref{Qrot}),
  based on a rigid rotor model, can also be used for symmetric and
  slightly asymmetric tops considering the rotational constants defined in the Principal Axis System~\citep{mcdowell1990,Herzberg,mangum2015}.

\subsection{Vibrational partition function}
\label{subsec-Qvib}

The vibrational partition function $Q_{\rm vib}(T)$, which only takes into
consideration the small amplitude vibrational modes, can be computed
using the harmonic approximation~\citep{Herzberg}:

\begin{equation}
\label{Qvib-harm}
Q_{\rm vib}^{\rm harm}(T)=\Pi_{\rm i=1}^{N_{\rm vib}} {1 \over 1 -
  e^{E^{({\rm vib})}_{\rm i}/k T}}  ~~,
\end{equation}

\noindent where $N_{\rm vib}$ and $E^{({\rm vib})}_{\rm i}$ represent the number of the small-amplitude vibrational modes and their fundamental energies, respectively.
This approximation is a good alternative to the direct sum expression
of the vibrational partition function for the ISM temperatures for the
following { reasons. In general, the energy values of the fundamental 
    vibrational states are known but only a handful of excited
  term values are determined. Furthermore, the limited number of known
  excited terms is not considered, and as a consequence the
  exponential terms have 
  larger $E^{({\rm vib})}_{\rm i}$ with respect to the typical ISM
  temperatures (100-200~K). More specifically, the terms with large
  fractions of ${E^{({\rm vib})}_{\rm i} \over k T}$ can be neglected
  in the calculation of the vibrational partition function. Finally,
  the convergence of the vibrational partition function can be
  generally reached only using the lowest vibrational energies. 
}
  
However, two comments on Eq.(\ref{Qvib-harm}) need to be added. First, it is known that the vibrational anharmonicity affects,
in general, the vibrational energy 
structure of floppy or semi-floppy molecules, mainly the excited
vibrational levels, by decreasing the vibrational term values
$E^{({\rm vib})}_{\rm i}$
with respect to their harmonic values. Therefore, by including the
anharmonicity, the direct sum of the vibrational partition function
increases~\citep{zheng2012,skouteris2016}. Second, molecules can undergo strong vibrational resonances
such as the Fermi and Darling-Dennison interactions~\citep{Herzberg},
which perturb the energy levels (sometimes by hundreds of
  cm$^{-1}$). These perturbation effects are not reflected in the
fundamental energies used in Eq.~(\ref{Qvib-harm}).

Therefore, it is worthwhile to test whether the harmonic expression
for the vibrational partition function of Eq.~(\ref{Qvib-harm}) is a good
approximation of the direct sum, as \citet{furtenbacher2016} did
  in a comprehensive study of the rovibrational partition function for
  water molecules. In order to assess the deviation of the harmonic partition
function with respect to the direct sum, we therefore chose the hydrogen sulfide molecule H$_2$S because: (i) this
  molecule has strong Fermi and Darling-Dennison interactions and
  large anharmonic effects in the vibrational degrees of freedom; and (ii) 
 a huge amount of compiled experimental and predicted vibrational levels (204
  vibrational term values in total up to approximately 17000~cm$^{-1}$) are available~\citep{carvajal2015}. 
  Therefore, H$_2$S is a good molecule to test the
    effect of the anharmonicity on the partition
    function because concerning the term values of
      its vibrational levels we can increase the temperature high enough to compare the direct sum with the harmonic approximation.
    Taking into account the vibrational levels of H$_2$S we
  calculate the values of the vibrational partition function
  contribution by direct sum. At $T=1000$~K and
  $2000$~K the resulting values of the
  direct sum are $1.28309$ and $2.45873$,
  respectively, and the relative differences with the harmonic
  approximation calculated from Eq.(\ref{Qvib-harm}) are of $0.13$\%
  and $1.34$\%,  respectively. Therefore, we 
    expect that within the typical ISM 
  temperatures, which are lower than the two temperatures given above,
  the harmonic vibrational partition function 
expression can in general be considered as a good approximation.

\subsection{Torsional partition function}

In the particular case of molecules with one large-amplitude
torsional CH$_3$ top, the torsional contribution to the partition
function $Q_{\rm tor}(T)$ can be computed by the direct
  sum~\citep{favre2014}:

\begin{equation}
\label{Qtor}
Q_{\rm tor}^{{\rm v}_{\rm t}^{\rm max}}(T) = \sum_{{\rm v}_{\rm
    t}=0}^{{\rm v}_{\rm t}^{\rm max}} \left(e^{-{E^{({\rm tor})}({\rm v}_{\rm t},A)
    \over k T}} + e^{-{E^{({\rm tor})}({\rm v}_{\rm t},E) \over k T}} \right),
\end{equation} 

\noindent where $E^{({\rm tor})}({\rm v}_{\rm t},A)$ and $E^{({\rm
    tor})}({\rm v}_{\rm t},E)$ are the torsional
     substate energies for the $A$ (either
    $A_1$ or $A_2$) and $E$
    symmetries (according to the molecular symmetry
group ${\cal C}_{\rm 3v}(M)$), respectively, { relative to the}
torsional v$_{\rm t}=0$ ground
  state, that is, $E^{({\rm tor})}({\rm v}_{\rm t}=0,A)=0$~cm$^{-1}$.

Different
    approximations for the torsional partition sum can be considered in terms of the
    selection of the maximum torsional quantum number
    v$_{\rm t}^{\rm max}$. For the particular case of methyl
      formate isotopologs, the convergence is achieved within 1\%
      when v$_{\rm t}^{\rm max}=6$ for a
temperature range up to $T=$300~K \citep[see Tables 9 and 10
of][]{favre2014}. However, as the convergence of
  Eq.~(\ref{Qtor}) depends on the torsional term values, which are specific to each molecule, and
  on the temperature, it is recommended whenever possible to perform
  a convergence test of the values obtained for the partition function
  from the direct sum by simply increasing v$_{\rm t}$ and adding the
  corresponding energies in the partition function until
  a negligible change of its value occurs. 

For molecules with two or more large-amplitude
torsional CH$_3$ tops, the number of exponential terms in
Eq.~(\ref{Qtor}) should be extended to all torsional substates for a
given torsional quantum number v$_{\rm t}$.

\subsection{Nuclear spin statistical weights}
\label{subsec-gu}

 The nuclear spin statistical weight $g_{\rm ns}$ can be
 computed according to \citet{BJbook}. A complete description of
   the calculation can be found in this latter reference, and so we only
   summarize it here. The nuclear spin statistical weight is obtained
 taking into account that the sign of the complete 
internal wavefunction\footnote{The complete internal
   wavefunction is the eigenfunction of the complete Hamiltonian for
     the internal dynamics of the molecule, which includes the
     rovibronic, electron spin, and nuclear spin degrees of
     freedom~\citep{BJbook}.} either changes or is kept
   invariant under the odd molecular symmetry group permutations of
   identical nuclei with  half-integer spin (so-called 
   {\em fermions}) or with integer spin (named {\em bosons}),
   respectively. For both cases however, the
     parity of the complete internal wavefunction can be positive or
     negative under the symmetry group inversion action~\citep{BJbook}. 

   Since the complete internal wavefunction with
   symmetry $\Gamma_{\rm int}$ can be written as the product of the
   rovibronic function, with symmetry $\Gamma_{\rm rve}$, and the nuclear
   spin function, with symmetry $\Gamma_{\rm ns}$, the
   following relationship exists among the various symmetry representations:
    $\Gamma_{\rm int} \subset \Gamma_{\rm rve} \otimes \Gamma_{\rm ns}$.
   Therefore, the spin statistical weight  $g_{\rm ns}$ is the number of  complete
   internal wavefunctions of the allowed $\Gamma_{\rm int}$ symmetry that
   can originate from a given $\Gamma_{\rm rve}$~\citep{BJbook}.

In general, $g_{\rm ns}$ is split into the reduced nuclear
spin weight $g_{\rm I}$ and the $K$-level degeneracy $g_{\rm K}$~\citep{turner1991,mangum2015} as follows:

\begin{equation}\label{gns}
g_{\rm ns}=g_{\rm I} \, g_{\rm K}  ~~.
\end{equation}

A possible source of error could come from the fact that the left part of Eq.~(\ref{logInt}), 
$\ln \left[\frac{3 \, k \, W}{8 \, \pi^3 \, \nu \, S\mu^2 \, g_{\rm u}}
\right]$, contains the statistical weight of the upper
state, but at the same time the right part of Eq.~(\ref{logInt})
contains the partition function $Q(T)$, which also depends on the
statistical weight. Inconsistencies and
errors therefore occur when taking from the literature a value for
the partition function $Q(T)$ (where the statistical weight $g_{\rm ns}$ is calculated in one way), and
then another value for the upper level degeneracy $g_{\rm u}$
(linked to $g_{\rm I} \, g_{\rm K}$ through Eq.~(\ref{gu})) calculated in another way. Of
course, this source of error can easily be resolved if the values of
$g_{\rm ns}$  are reported apart
from the rest of the partition function, allowing the users to
carefully check the values of $g_{\rm ns}$ and $g_{\rm u}$ before considering the partition
function in Eq.~(\ref{logInt}).

\subsection{Interpolation and extrapolation}
\label{subsec-interpol-extrapol}

Whenever possible, the best procedure is to calculate the direct
  sum using Eq.~(\ref{rtvQ}) after making sure that the sum reaches convergence
  for the temperature under study. This of course may not be always
  possible since it requires a rather precise knowledge of the
  vibrational, rotational, and torsional (if large amplitude
    torsional motion(s)
  is involved) energies. Furthermore, depending on the temperatures studied, those
  energy levels may not be available through experimental transition
  measurements or even through calculated transitions. In those cases, 
 the partition function can alternatively be tabulated, within
a given temperature range, via fitted mathematical parameters using different power
expansions depending on the
temperature~\citep{irwin1981,fischer2002,barton2014}. These
power expansions, apart from being used as a simple way of storing
or computing the
rotational-vibrational-torsional partition functions, are useful to
interpolate and/or extrapolate  
the partition function values to the temperature of any ISM
region, and they are therefore often used. 
Nevertheless, the main advantage of the tabulation of the partition
  function is due to the fact that in general only a set of partition function
  values are reported for a set of temperatures (typically from 2.725~K up to
  500~K) and the direct sum cannot be obtained if the rovibrational energies are not available, as is usually the case.

In Sect.~\ref{subsec-molecules} we consider
  three molecular species as examples to analyze the 
  suitability of the most used procedures according to the ISM
  temperature range. This analysis led us to draw the
  conclusion that a noteworthy difference of around 10\% can be
  attributed to using one or another interpolation procedure. Two
    main interpolation procedures can be distinguished: the linear 
interpolation, 
which is obtained by simply taking two values of the partition function
at two successive 
temperatures and tracing a straight line between them (which of course
is strictly valid only between the two points), and the nonlinear
interpolation, which uses the power expansions of the 
  rotational-vibrational-torsional partition function
in terms of temperature.

As far as the nonlinear power expansions in the temperature are
  concerned, two kinds of methods are found in the literature. One
  method was considered by \citet{fischer2002}, who used a fourth-order
polynomial of $Q(T)$ in terms of temperature $T,$ and is valid for a temperature
range from 70 to 300~K: 

\begin{equation}\label{Qpolynomial}
Q(T)= \sum_{{\rm n}=0}^6 A_{\rm n} \,T^{\rm n} .
\end{equation}

In the present study, a polynomial extended
up to sixth order is considered, where seven parameters are fitted
from the computed partition function values. In the second
  method, a nonlinear fit of a power 
expansion of $\log_{10}(Q)$ in terms of the $\log_{10}(T)$ is also
found in the literature~\citep{irwin1981,barton2014}, as follows:

\begin{equation}\label{logQpolynomial}
\log_{10} Q(T) = \sum_{{\rm n}=0}^6 \, a_{\rm n} \, (\log_{10} \,
T)^{\rm n} ,
\end{equation}

\noindent where seven parameters are also fitted.

It is important to note that in comparison to the linear interpolation, Eqs.~\ref{Qpolynomial} and \ref{logQpolynomial} can be successfully used to
  interpolate the partition function along the whole temperature
  interval for which they were fitted, and in addition to extrapolate
  them to further values of the temperature interval used in the fitting.

Furthermore, we note that one can also use the rigid rotor approximation of~\cite{Herzberg} for symmetric and 
asymmetric tops, respectively, to derive the rotational partition
function at any temperature $T$, using the value of the partition
function at $T$=300~K as a benchmark:

\begin{equation}\label{Qextrapolation}
Q(T)=Q(300) \, \left({T \over 300}\right)^{1.5} .
\end{equation}

A  quantitative example of the
  extrapolation function (\ref{Qextrapolation}) is given in Sect.~\ref{subsec-MF}.

\subsection{Example of  molecular partition function calculation}
\label{subsec-molecules}

In this section we discuss the calculation of the partition
functions for the three following molecules that represent different
geometries: isocyanic acid (HNCO), which is a quasilinear molecule,
methyl cyanide (CH$_3$CN), a symmetric top molecule, and methyl
formate (HCOOCH$_3$), an asymmetric top molecule which in addition has
a large-amplitude internal rotor from CH$_3$ top. The nuclear spin statistics and the nonlinear polynomial expansions of the partition functions are discussed in Appendices~\ref{appendix-nuclear-spin} and \ref{appendix-non-linear-polyn}, respectively.

\subsubsection{Isocyanic acid (HNCO)}
\label{subsec-HNCO} 

Isocyanic acid is a slightly asymmetric prolate rotor~\citep{niedenhoff1995}.
The shortage of spectral data from the different vibrational bands
does not allow us to obtain the partition function  straightforwardly
from Eq.~(\ref{rtvQ}) as far as the vibrational energies are concerned. Nevertheless, there are enough rotational
levels that can be deduced from measured
transitions~\citep{kukolich1971,hocking1975,niedenhoff1995,lapinov2007}.
Moreover, fundamental vibrational term values were already recorded in the gas
phase~\citep{east1993} so one can compute the
partition function using Eq.~(\ref{rtvQapprox}). The torsional
term is not considered at all because this motion does not occur
  in HNCO.

In Table~\ref{tab-Qrot-HNCO} we compare
the values of the rotational partition function for different temperatures ($Q_{\rm rot}^{\rm
  approx}$ in the second column) using the approximation for slightly
asymmetric tops~\citep{mcdowell1990}
with the value of the rotational partition
function computed as a direct sum of the predicted rotational energy levels up to
  J=135 and K$_a$=30~\citep{lapinov2007} using Eq.~(\ref{Qrot}) without considering
the nuclear spin statistical weight ($Q_{\rm
    rot}(\mbox{\rm direct sum})$ in the third
column). It is clear from
Table~\ref{tab-Qrot-HNCO} that the approximation of
\citet{mcdowell1990} or similar approximations are far from
being satisfactory at temperatures lower than 20~K.

As far as the vibrational contribution of the partition function is
concerned, there are not enough observed or computed vibrational
energies available to compute the direct sum. One needs to use the
harmonic approximation of the vibrational partition function from
Eq.~(\ref{Qvib-harm}). In Table~\ref{tab-PartF}, the values of the harmonic
  approximation of the vibrational partition function
  $Q_{\rm vib}^{\rm harm}$ from Eq.~(\ref{Qvib-harm}) are multiplied
  by the rotational partition function to obtain
the rotation-vibration partition function $Q_{\rm
      rv}(\mbox{\rm Present work})=Q_{\rm
    rot}(\mbox{\rm direct sum}) \cdot Q_{\rm vib}^{\rm harm}$ for the isocyanic acid
(HNCO) molecule in the temperature 
range from 2.725~ to 500~K. The values of the
  rovibrational contribution from the present study are compared with those
coming from the CDMS catalog $Q$(CDMS) which considers only
the rotational partition function as a direct sum and no
  vibrational contribution. It is notable that here the
vibrational contribution is becoming significant at temperatures
higher than 225~K. Therefore, if the molecule emits in the ISM
object at a temperature greater 
than 200~K, it is advisable to consider the vibrational
contribution in the calculation of the molecular partition function.

The relative differences of $Q_{\rm
     rv}(\mbox{\rm Present work})$ with respect to $Q$(CDMS) are in
   general larger than the uncertainties of $Q_{\rm
     rv}(\mbox{\rm Present work})$. An upward estimate of its
   uncertainty has been provided in
   Table~\ref{tab-PartF} considering large uncertainties  of 100~MHz
   and 1~cm$^{-1}$ for the rotational energies and the vibrational fundamentals, respectively. We have also computed the uncertainty of the $Q_{\rm
     rv}(\mbox{\rm Present work})$ values using the uncertainties for
   each rotational energy level~\citep{pickett1991}, but these are even smaller.

{ In Table D.1 (available at the CDS),} the computed rotational, vibrational, and
  rovibrational partition function $Q_{\rm
     rv}(\mbox{\rm Present work})$ for isocyanic acid (HNCO)
  are also given up to  T=500~K in intervals of 1~K.

\subsubsection{Methyl cyanide (CH$_3$CN)}
\label{subsec-CH3CN}

Methyl cyanide, also named acetonitrile, is a symmetric rotor very
close to the prolate limit~\citep{muller2015}. As for isocyanic acid, the partition
function can be calculated through Eq.~(\ref{rtvQapprox}) after
computing the rotational partition function and the vibrational
contribution separately. There is no torsional contribution. The rotational
energy data were taken { from \citet{kukolich1973}, \citet{boucher1977},
\citet{kukolich1982}, \citet{cazzoli2006}, and \citet{muller2009}
while} the vibrational fundamental frequencies come from \citet{rinsland2008}.

In Table~\ref{tab-Qrot-CH3CN}, the classical approximation of the
rotational partition function (using \citet{mcdowell1990}) is
compared for CH$_3$CN to the direct sum
expression for the rotational partition function (Eq.~\ref{Qrot}) but omitting the nuclear spin weight. The
classical approximation has been  
computed using the rotational constants from \citet{muller2015} while
the direct sum has been computed for all the rotational energy
  levels up to J=99~\citep{muller2009}. It
can be noted that the 
classical approximation for the rotational partition function
contribution is not accurate enough  for temperatures lower than
$5$~K. Hence, in the present study, the rovibrational partition 
function has been calculated as the product of the direct sum of the
rotational contribution and the harmonic vibrational approximation
(see Table~\ref{tab-PartF}). 

In Table~\ref{tab-PartF}, the values of the rovibrational partition function $Q_{\rm rv}(\mbox{\rm Present work})=Q_{\rm rot}(\mbox{\rm direct sum}) \cdot Q_{\rm vib}^{\rm harm}$
are given from 2.725 to 500~K (column 3) and are compared with the
rovibrational partition function provided by the CDMS
{ catalog \citep[column 4; ][]{endres2016};} the vibrational
  partition function contribution of this catalog only takes into consideration the vibrational fundamentals up to about 1200~cm$^{-1}$. For
comparison purposes, the nuclear spin statistical weight has been
suppressed in both contributions. It can be 
noted in Table~\ref{tab-PartF} that they disagree at temperatures
  lower than 20~K and larger than 300~K. At temperatures below 20~K
  the discrepancy, which could only stem from the  
rotational partition function, 
could come from an extra contribution overlooked in the instructions
since the
completeness of the rotational partition function calculated in the
present study has also been proven by the good agreement obtained with the
rotational contribution computed by
\citet{rinsland2008}. For the highest temperatures, the difference between these
two calculations (CDMS vs. this work) only comes from the vibrational
contribution of the partition function, which starts to become significant
for temperatures above 300~K.

The uncertainties of $Q_{\rm rv}(\mbox{\rm Present work})$ are
   also given in Table~\ref{tab-PartF}. They are rather small
   despite them being estimated upwards considering large rotational
   energy uncertainties of 100~MHz and of 1~cm$^{-1}$ for the
   vibrational fundamentals. In a temperature range above 20~K, the
   uncertainties of $Q_{\rm rv}(\mbox{\rm Present work})$ are
   significantly smaller than the relative differences with respect to
   $Q$(CDMS).

{  In Table D.2 (available at the CDS),} the rotational, vibrational, and
  rovibrational partition functions calculated in the present study for
  methyl cyanide (CH$_3$CN) are provided in intervals of 1~K up to T=500~K.

\subsubsection{Methyl formate  (HCOOCH$_3$)}
\label{subsec-MF}

Methyl formate is an asymmetric near-prolate rotor which
has a large-amplitude motion due to the torsion of the methyl
group. As a consequence, each rotational line is split into a
doublet characterized by the symmetry labels A$_1$ or A$_2$ and
E~\citep{carvajal2007}, respectively. Therefore, for this molecular species, the torsional contribution has to be taken into account in the approximated expression of
the partition function~(Eq. \ref{rtvQapprox}). Other species of
  astrophysical interest undergoing a large-amplitude motion of the
  methyl group CH$_3$ are, for example, methanol and acetaldehyde (\citealt{voronkov2002,martin2006,xu2008,wang2011,xu2014,slocum2015,pearson2015,kleiner1996,kleiner1999,kleiner2008,elkeurti2010,smirnov2014,margules2015,codella2016,zaleski2017}).

A number of spectroscopic studies have been carried out
for the identification of the isotopologs of methyl formate in the
ISM~\citep{carvajal2009,margules2010,takano2012,tercero2012,favre2011,favre2014,haykal2014}. The
spectral analyses were focusing on the rotational features of these molecular
species mainly in the ground and first torsional excited states~\citep{curl1959,brown1975,bauder1979,demaison1983,plummer1984,plummer1986,oesterling1999,karakawa2001,odashima2003,ogata2004,willaert2006,carvajal2007,maeda2008a,ilyushin2009,carvajal2010,tudorie2012,duan2015}
while the other vibrational bands were in general not
considered except in a few cases~\citep{maeda2008b,kobayashi2013,kobayashi2018}.

A significant achievement was when the lowest
torsional-rotational transitions for internal rotors could be reproduced
and predicted using the so-called {\it Global Approach}, where the torsion-rotation
Hamiltonian matrix (containing higher-order terms describing the
coupling between rotation and torsion) is set up in an extended basis
set containing the nine lowest torsional v$_{\rm t}$ states. By
fitting v$_{\rm t}=0$ and $1$ bands within
experimental-accuracy laboratory data, \citet{tudorie2012} provided a set of 53
parameters with a weighted (unitless) root-mean-squared deviation of 0.67. Even though the
experimental data only included v$_{\rm t}=0$ and $1$, the fitted parameters
contain implicitly all the interactions involving v$_{\rm t} < 9$. Predicted
energy levels higher than v$_{\rm t}=1$ can therefore in principle be computed using those parameters
 but care has to be applied as the extrapolation to
higher torsional levels lacks accuracy as they are not based on
experimentally fitted parameters.  

Therefore, a rotational-torsional-vibrational partition function
can only be computed in an approximated way. The vibrational
contribution was obtained via the harmonic approximation formula
(Eq. \ref{Qvib-harm}) by using experimental vibrational fundamental
frequencies provided by \citet{chao1986}, while the 
  rotational and torsional contributions were obtained by
  performing a direct sum on the predicted energies.

\paragraph{Comparison of rotational and rotational-vibrational-torsional partition
  functions.}

It is worth noting that the most extensive
  rotational-torsional partition function calculation of methyl formate was provided by
  \citet{tudorie2012} although this calculation was restricted to
  temperatures from $100$~ to $300$~K in steps of
  $10$~K. This partition function was computed as a direct sum of the
  predicted rotational-torsional energies calculated using
  the Hamiltonian parameters obtained in a fit of observed transition
  lines in the ground 
  and first excited torsional v$_{\rm t}=0$ and $1$ states. The
  vibrational partition function was not provided in
  \citet{tudorie2012}. 

In 2014, \citet{favre2014} provided
rotational-torsional-vibrational partition function values for methyl
formate in the temperature range from 9.375~ to 300~K after a
comprehensive convergence study was carried out. More specifically, the partition
function was computed using Eq.~(\ref{rtvQapprox}). The
rotational contribution was calculated as a direct sum (Eq. \ref{Qrot}) for
the rotational-torsional energies with torsional quantum number 
v$_{\rm t}=0$ and $A_1$ or $A_2$ symmetries up to $J=79$. 
The torsional (corresponding to the large-amplitude motion) contribution was
calculated as a direct sum by \citet{favre2014} using Eq.~(\ref{Qtor})
up to v$_{\rm t}^{\rm max}=6$ and
 the following torsional energies: i) for the torsional levels
v$_{\rm t}=0$ to v$_{\rm t}=2$, from the measured and predicted transitions
coming from \citet{ilyushin2009}, ii) for the
levels v$_{\rm t}=3$ to v$_{\rm t}=4$, {\it ab initio} torsional term values were
used~\citep{senent2005}, and iii) for the levels v$_{\rm t}=5$ to
v$_{\rm t}=6$ a
harmonic estimation from those torsional levels were used. The vibrational
contribution of the small-amplitude vibrations was obtained using the harmonic
approximation~(Eq.~\ref{Qvib-harm}).

In the present study, the calculation of
the rotational-torsional-vibrational partition function is
revisited. The values obtained between 9.375~ and 300~K by
  \citet{favre2014} are extended here from 2.725~ to
500~K. Despite the fact that the partition function of \citet{favre2014} provided
good agreement with respect to that given by \citet{tudorie2012}, it
is expected that our present study gives more accurate partition
function values mainly because of the increase of the maximum
  torsional quantum number v$_{\rm t}^{\rm max}$ in Eq.~(\ref{Qtor}), as
  described below.  

\citet{favre2014} verified that, for methyl formate, the classical
approximation~\citep{Herzberg} of the rotational partition function provided
satisfactory results compared to its direct
sum expression for rotational energies up to $J=79$ and for a temperature
range from 9.375~ to 300~K. Table~\ref{tab-Qrot-MF} shows
that the classical approximation for methyl formate is fairly good even for low
temperatures around 2.725~K.
In fact, the
comparison of the values provided by the direct sum and the classical
expression represents a good validation of the convergence reached by
the rotational contribution of the partition function in the
  temperature range up to 500~K.

To extend the torsional contribution of the partition function
at temperatures from 300~ to 500~K, we used Eq.~(\ref{Qtor}) with torsional states up
to v$_{\rm t}^{\rm max}=10$. The torsional energies up to
v$_{\rm t}=4$ were predicted from our internal rotation software {\it
  BELGI}~\citep{kleiner2010} using the Hamiltonian parameters
experimentally fitted by \citet{tudorie2012}. The torsional energy
levels from v$_{\rm t}=5$ to v$_{\rm t}=10$ were estimated with a harmonic
approximation. 
Although it is known that the harmonic energy estimate
provides a smaller value for the torsional partition function, in comparison with
the one obtained when all the experimental torsional levels are
available, this approximation is still justified for methyl
formate and provides a minimum threshold for the torsional
  partition function values at high temperatures. Indeed, as
mentioned above, the higher excited torsional energies can only
be extrapolated from the experimentally deduced 
parameters of the Hamiltonian from observed transitions within v$_{\rm t}=0$
and $1$.

In Table~\ref{tab-Qtor-Qvib-MF}, the
torsional partition function calculated as a direct sum in the present study are given for
various
v$_{\rm t}^{\rm max}$ equal to $4$, $6$, $8,$ and $10$.
We can make one comment about the fact that \citet{favre2014}
presented the cutoff at v$_{\rm t}^{\rm max}=6$ for temperatures
up to 300~K. In fact, the
deviation between v$_{\rm t}^{\rm max}=6$ and v$_{\rm t}^{\rm max}=8$ 
approximations is smaller than 1\%, and between
  v$_{\rm t}^{\rm max}=8$ and v$_{\rm t}^{\rm max}=10$ it is of 0.22\% at
  $T$=300~K. Therefore, the convergence at $T$=300~K can be considered as accomplished. However,
at $T$=500~K, the difference in extending the calculation of the
torsional partition function from  v$_{\rm t}^{\rm max}=6$ to
v$_{\rm t}^{\rm max}=8$ is still 3.6\%. Extending the calculation from
v$_{\rm t}^{\rm max}=8$ up to v$_{\rm t}^{\rm max}=10$ torsional levels
produces a difference in the torsional partition function of 1.6\%
at $T$=500~K.
For completeness, Table ~\ref{tab-Qtor-Qvib-MF} also gives the harmonic approximation
of the vibrational (for the modes other than the torsional mode)
contribution of the partition function.

The rotational-torsion-vibrational partition function values
obtained in the present study as the product of the rotational, torsional, and vibrational
contributions (Eq.~\ref{rtvQapprox}) are given in the third
  column of Table~\ref{tab-PartF-MF}
for a temperature range from 2.725~ to 500~K and are compared to
  the values of \citet{favre2014} up to 300~K. The differences between
  the present study and the values of \citet{favre2014} are very small
  but increase with temperature. A conservative estimate of the
   uncertainty of $Q_{\rm rvt}(\mbox{\rm Present work})$ has also been
   provided, considering the rotational energy uncertainties
of 100~MHz, the uncertainties of the vibrational fundamentals as
1~cm$^{-1}$ and the uncertainties of the torsional energies as
0.01~cm$^{-1}$ for v$_{\rm t}$=0 and 1, 1~cm$^{-1}$ for v$_{\rm t}$=2 states,
5~cm$^{-1}$ for v$_{\rm t}$=3 and 4, and 20~cm$^{-1}$ for the
remainder (v$_{\rm t}$=5-10).
Despite the upward estimate of the uncertainties, they are smaller
than 0.5\% of the partition function value.

Finally, we note that for $T$=500~K we used the
rotational partition function value given by the
classical approximation~\citep{Herzberg}.
This is justified because the
  relative difference between the direct sum and the approximated
  rotational partition function tends to decrease with temperature
  (see Table~\ref{tab-Qrot-MF}). 
 Indeed, for temperatures higher than 330~K, the direct sum expression does
 not reach the convergence  if the sum
is limited to rotational energy levels up to $J=79$, which is the
  limit where we measured and fit the experimental
  data~\citep{kleiner2010}. 

{  In Table D.3 (available at the CDS),} the values of the rotational, torsional, vibrational, and
rotational-torsional-vibrational partition function computed in the
present study for methyl formate are provided up to T=500 K using an
interval of 1~K.

In Fig.~\ref{fig:Q-MF}, the partition function for the
  main isotopolog of methyl formate computed in the present study
  and the one published in the JPL catalog are compared by
  considering $g_{\rm ns}=1$ in both cases. The partition function of the
  present study was extended up to $T$=500~K and is plotted as full red
  circles. The JPL partition function is displayed in full green
  squares. The JPL values were computed as a direct sum
(\ref{rtvQ}) of all the
predicted rotational-torsional energies with rotational angular
momentum up to $J=109$ and including the ground and first torsional
levels v$_{\rm t}=0$ and $1$~\citep{ilyushin2009}. The JPL
values~\citep{pickett1998} were given in the 
typical set of temperatures reported in the
astronomical catalogs, that is, $T$=9.375, 18.75, 37.50, 75.0,
150.0, 225.0, and 300.0~K.

Figure~\ref{fig:Q-MF} shows that the partition function deviations
  between the present study and the JPL catalog values start to be important at
  $T\ge$150~K. Indeed, at $T$=150~K, they differ by about 20\% while at
  $T$=225~K and $T$=300~K the deviations are $\sim$43\% and $\sim$61\%,
  respectively. These differences are due to the fact that the JPL
  partition function does not include the vibrational contribution and
  takes into account the torsional contribution only up to 
  v$_{\rm t}=0$ and $1$, whereas our calculation includes torsional levels up to
  v$_{\rm t}=10$ as well as the vibrational contribution.

\paragraph{Comparison of partition functions from nonlinear
  polynomial expansions and direct sums.}

In order to better assess the rotational-vibrational-torsional
  partition function obtained at the estimated temperature of the ISM
  survey from the linear and nonlinear interpolation procedures (see Appendix~\ref{appendix-non-linear-polyn}),
we plot the values of $Q$ against the temperature in Fig.~\ref{fig:Q-MF}. The 
results of the { linear interpolation (traced as a straight line by considering  the sets of two adjacent partition function points) and those from the nonlinear interpolation fitting (in this
case the sixth-order polynomial expansion, from Eq.~\ref{Qpolynomial})
are also displayed.}
It is immediately apparent that the computed
partition function values show a better agreement with  the nonlinear interpolation (third column of Table~\ref{tab-PartF-MF})
than the linearly interpolated 
partition function values. In fact, the linear interpolation provides
overestimated values for the partition function (which become even larger for
higher temperatures). Therefore, using a linear interpolation of the
partition function will lead to overestimation of the molecular column
density according to Eqs.~(\ref{logInt}) and~(\ref{logInt2}).

\begin{figure*}
\centering
    \includegraphics[width=16cm]{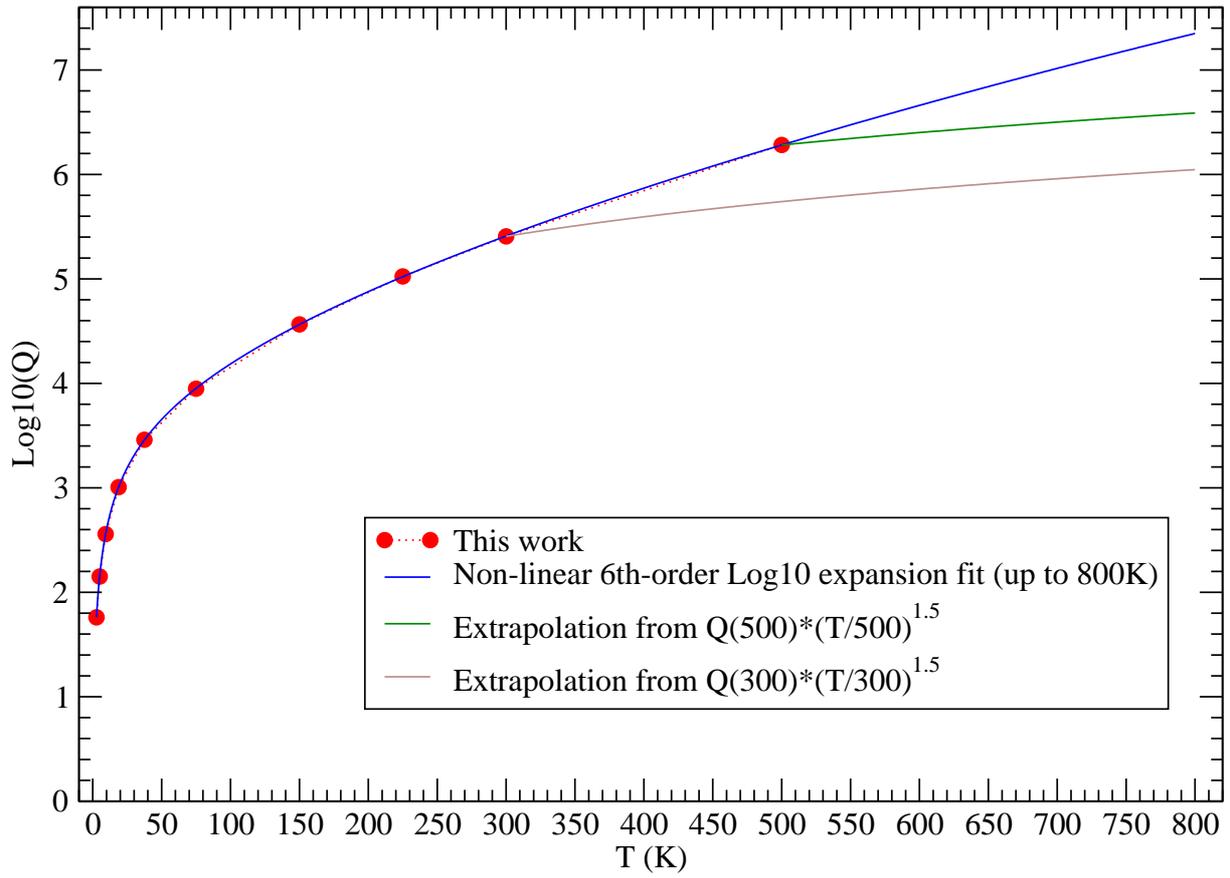}
\caption{Logarithm of Q in base 10 is provided according to
  calculation of the present study (full red circles). The nonlinear
  sixth-order decimal logarithm polynomial curve fitting (blue line) is provided up to
  $T$=800~K. This result is compared with the
  extrapolated values obtained using the partition function at
  $T$=300~K  as benchmark  (brown line) (see
  Eq.~(\ref{Qextrapolation}). In addition, it has been included the
  extrapolated values computed with Eq.~(\ref{Qextrapolation}) but
  using $T$=500~K as benchmark (green line).}
\label{fig:log10Q-MF}
\end{figure*}

Figure~\ref{fig:log10Q-MF} shows the $\log_{10}(Q)$ values
computed in the present study with respect to the temperature from $T$=2.725~K up to
$500$~K (red full circles). The nonlinear curve fit, displayed in
this figure as a blue continuous line, is the sixth-order
decimal logarithm polynomial
  expansion (Eq.~\ref{logQpolynomial}).  
The latter was used to extrapolate the partition function to
temperatures from 500~ to 800~K. It is immediately apparent that
up to the temperature of 800~K, the polynomial expansion follows the same trend as the
lower temperatures partition function. In addition, the values that have been extrapolated via the polynomial expansions are consistent 
 with the present partition function values computed using
   the direct sum between 2.725~ and 500~K.

Finally, the extrapolation carried out with the sixth-order
decimal logarithm polynomial expansion
  (Eq.~\ref{logQpolynomial}) are compared with the
extrapolated values obtained using 
 the partition function at $T$=300~K (brown line) and $T$=500~K  (green
 line) using the expression of Eq.~(\ref{Qextrapolation}). 
 This leads us to conclude that for methyl formate the nonlinear fitting
expression is more suitable in the extrapolation
  procedure of the partition function values.

\section{A test case: observations of HNCO, CH$_3$CN, and HCOOCH$_3$
  towards the star forming region NGC7538--IRS1}
\label{sec-astro}

\subsection{Observations}
Observations of NGC7538--IRS1 were carried out with the IRAM 30 m
telescope at Pico Veleta in Spain on 2013 December 5, 6, and 10. The
observations were performed in the ``position switching'' mode, using
[$-$600$^{\prime\prime}$,0 $^{\prime\prime}$]  as reference for the
OFF position, toward a single pointing
($\rm\alpha_{J2000}$=23$\rm^{h}13^{m}45.\!\!^{\mathrm s}5$,
$\rm\delta_{J2000}$= $+$61$^\circ28^\prime12.\!\!^{\prime\prime}$0). 
The E230 EMIR receiver was used in connection with  the 200~kHz Fourier transform spectrometer (FTS) backend in the frequency ranges 
212.6--220.4~GHz, 228.3--236.0~GHz, 243.5--251.3~GHz,
251.5--259.3~GHz, 259.3--267.0~GHz, and 267.2--274.9~GHz. The
half-power beam size is 10$^{\prime\prime}$ for observations at
250~GHz. The $v$$_{\rm LSR}$ was $-$57~km~s$^{-1}$ and the spectral
resolution is about 1~km~s$^{-1}$. The resulting data harbored
standing wave together with spurs that were removed during the
reduction (using the fast Fourier transform for the standing
wave).

\begin{figure*}
\centering
    \includegraphics[width=22cm,angle=270]{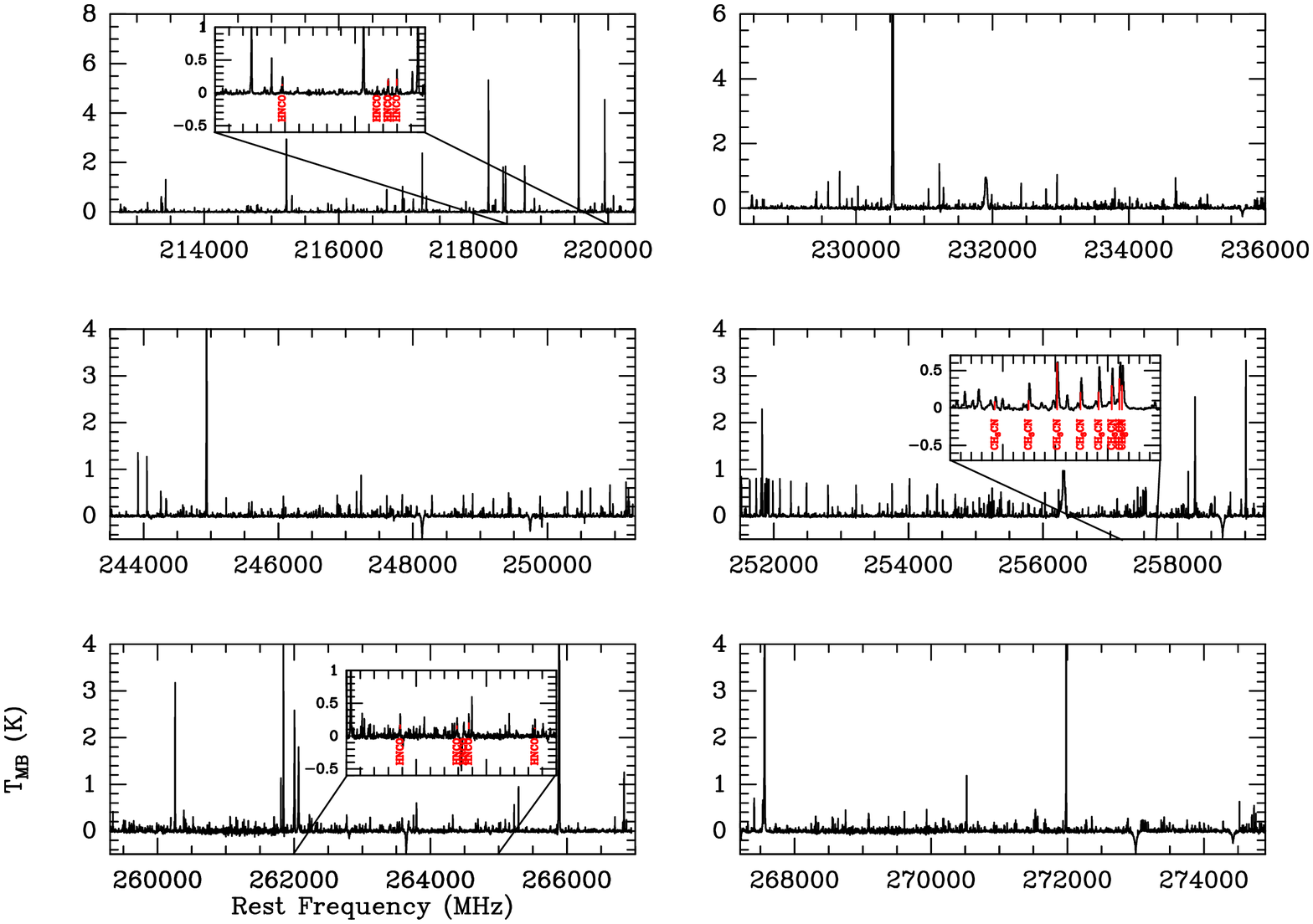}
\caption{NGC7538--IRS1 spectra as observed with the IRAM-30m telescope. Line assignment for CH$_3$CN and HNCO is shown in red. }
\label{fig:spectra}
\end{figure*}

\begin{figure*}
\centering
    \includegraphics[width=22cm,angle=270]{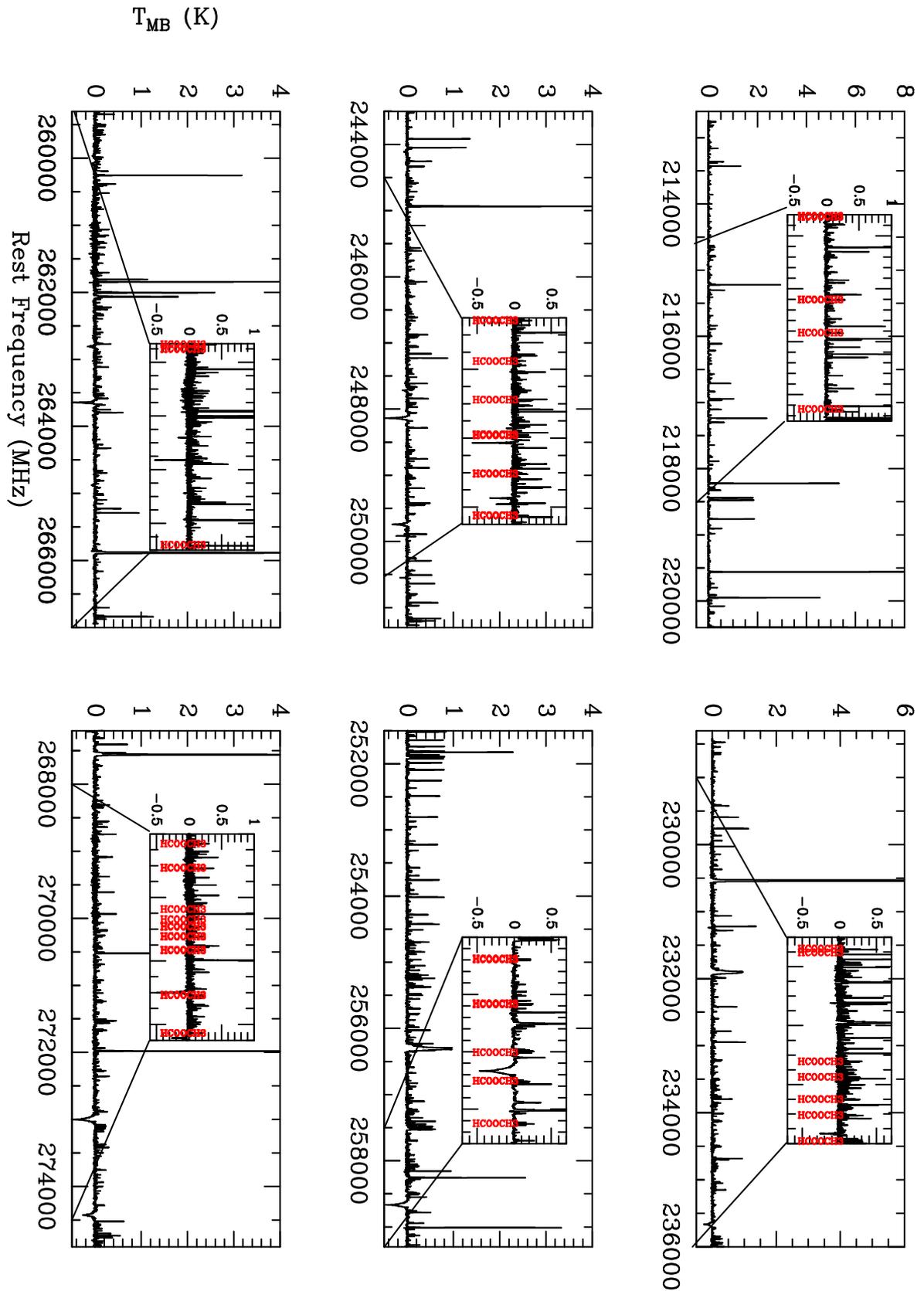}
\caption{ NGC7538--IRS1 spectra as observed with the IRAM-30m telescope. Line assignment for HCOOCH$_3$ is shown in red. }
\label{fig:spectra2}
\end{figure*}

Figures~\ref{fig:spectra} and \ref{fig:spectra2} display the NGC7538--IRS1 spectra observed
with the IRAM 30 m telescope in main beam temperature units
(T$\rm_{MB}$) that can be obtained with the following { expression,} 
\begin{equation}\label{tatotmb}
T\rm_{MB}= {\eta\rm_{f} \over \eta\rm_{MB}} \times T\rm_{A}^{*} ~~,
\end{equation}
where $T\rm_{A}^{*}$ is the antenna temperature, $\eta\rm_{f}$ the
forward efficiency, and $\eta\rm_{MB}$ the main beam efficiency. In
this paper, we used $\eta\rm_{f}$ = 94, 92, and 87 and
$\eta\rm_{MB}$= 63, 59, and 49 at 210, 230, and 274~GHz,
respectively\footnote{see
  http://www.iram.es/IRAMES/mainWiki/Iram30mEfficiencies}.

\subsection{Molecular frequencies}

For HNCO we used the spectroscopic data parameters from \citet{kukolich1971},
\citet{hocking1975}, \citet{niedenhoff1995}, and
\citet{lapinov2007}. For CH$_3$CN we used the data from
\citet{muller2009}, \citet{cazzoli2006}, { \citet{kukolich1973}, \citet{kukolich1982},} \citet{boucher1977},
 \citet{anttila1993}, and \citet{gadhi1995}, which are available
 at the Cologne Database for Molecular Spectroscopy catalog 
\citep[CDMS,][]{Muller2005}. Regarding HCOOCH$_3$, we used spectroscopic parameters from
\citet{ilyushin2009}, \citet{brown1975}, \citet{bauder1979},
\citet{demaison1983}, \citet{plummer1984}, \citet{plummer1986},
\citet{oesterling1999}, \citet{karakawa2001}, \citet{odashima2003},
\citet{ogata2004}, \citet{carvajal2007}, \citet{maeda2008a},
\citet{maeda2008b}, and \citet{curl1959} available in the JPL catalog.

In particular, we searched for transitions up to $E_\mathrm{u}$ $\simeq$
460~K with an Einstein coefficient, $A_\mathrm{ij}$ larger than
1$\times$10$^{-4}$~s$^{-1}$ for HNCO and  $A_\mathrm{ij}$ $\ge$
1$\times$10$^{-3}$~s$^{-1}$ for  CH$_3$CN. Regarding HCOOCH$_3$, we
searched for transitions up to $E_\mathrm{u}$ $\simeq$ 310~K with $A_\mathrm{ij}$ larger than
2.9$\times$10$^{-5}$~s$^{-1}$. The spectroscopic parameters are listed
in Appendix~\ref{appendix-observational}.

\subsection{Observational line parameters}

The analysis presented below hinges upon the assumptions that \textit{i)} LTE is
reached (see Section~\ref{sec-rotdiag}), \textit{ii)} all the lines
are optically thin, \textit{iii)} the molecular emission arises
within the same source size, and \textit{iv)} the source size is equal to the beam size.

Tables~\ref{tabhnco}, \ref{tabch3cn},  and \ref{tabhcooch3} in
  Appendix~\ref{appendix-observational} summarize the following
observational line parameters for the targeted molecules derived from Gaussian fits: the LSR velocity v$_{\rm LSR}$
(km~s$^{-1}$), the line width at half intensity $\Delta$v$_{\rm LSR}$
(km~s$^{-1}$), the brightness temperature $T_{\rm B}$ (K), the integrated
line intensity $W$ (K~km~s$^{-1}$), and the column densities of the upper
state level of the transition with respect to the upper state
degeneracy $N_{\rm u}/g_{\rm u}$ (cm$^{-2}$). In those tables we also
  indicated for HNCO and CH$_3$CN the transitions that are
clearly detected, partially blended (i.e., if the emission is partially
contaminated by the emission from another molecule), or completely blended. Regarding HCOOCH$_3$, we only selected the
clearly detected transitions.

\subsection{Temperature}\label{sec-rottemp}

The temperature diagram method (based on Eq.~\ref{logInt} or
Eq.~\ref{logInt2} together with the observational line parameters)
allows us to derive the temperature at which the targeted
molecules are emitting in the observed ISM region. Figures~\ref{fig:NvsQ_CH3CN}, \ref{fig:NvsQ_HNCO}, and
\ref{fig:NvsQ_HCOOCH3} show the temperature diagrams derived from the
analysis for the CH$_3$CN, HNCO, and HCOOCH$_3$ line emission, respectively. 
The  resulting temperatures, $T$, are (350$\pm$31)~K for
CH$_3$CN, (349$\pm$77)~K for HNCO, and (182$\pm$44)~K for HCOOCH$_3$.
The derived temperatures are consistent, for HNCO and HCOOCH$_3$, with
the previous study of this source by \citet{Bisschop:2007}.

\subsection{Column densities}
The total column density, $N$, of a molecule can be obtained from
Eq.~(\ref{logInt} or \ref{logInt2})
using the partition
function, $Q(T)$, at the derived excitation temperature.  
In this paper, we aim to quantify the impact of the partition function
on the derived total column density. In that context, we used
different approximations of the partition function to derive the
total beam-averaged\footnote{In this study, we assume a source equal
  to that of the beam.} column density for CH$_3$CN, HNCO, and
HCOOCH$_3$. The results are given in Table~\ref{tab-ColumnDens}
for the rotational-vibrational(-torsional) partition functions
  obtained in the present study (third and seventh columns of
    Table~\ref{tab-PartF} for CH$_3$CN and HNCO, respectively, and third
    column of Table~\ref{tab-PartF-MF} for HCOOCH$_3$) and for the
approximations reported in the JPL or CDMS catalogs. In addition, we compared a linear and a
nonlinear interpolation carried out at the temperature of NGC-7538 survey.
 The more reliable estimate of the total column density is the one that is 
 calculated with the partition functions computed in the present study and given in the column 8 of Table~\ref{tab-ColumnDens}. It can be noted that the nonlinear
   interpolation of the partition function given in the present study
   (see Appendix~\ref{appendix-non-linear-polyn}) provides the same
   result for the total 
   column densities. The relative difference between the different methods (see Table~\ref{tab-ColumnDens}) is discussed in Section~\ref{sec-discussion}.

\section{Discussion: implications of the partition function on the ISM molecular column densities}
\label{sec-discussion}

In this section we show how the use of different approximations
for the rovibrational partition function together with some
interpolation and/or extrapolation procedures affects  estimations of the interstellar
molecular column density. 
In that context, Table~\ref{tab-ColumnDens} gives the ISM column
densities for CH$_3$CN, HNCO, and HCOOCH$_3$ derived from two different
approximations of the respective partition functions. The
calculations make use of the  excitation temperatures given in
Section~\ref{sec-rottemp} along with the partition functions 
obtained in the present study (see Table~\ref{tab-PartF} for
  CH$_3$CN and HNCO and Table~\ref{tab-PartF-MF} for HCOOCH$_3$)
and in the CDMS and/or JPL catalogs.

\begin{figure}
\centering
    \includegraphics[width=\columnwidth]{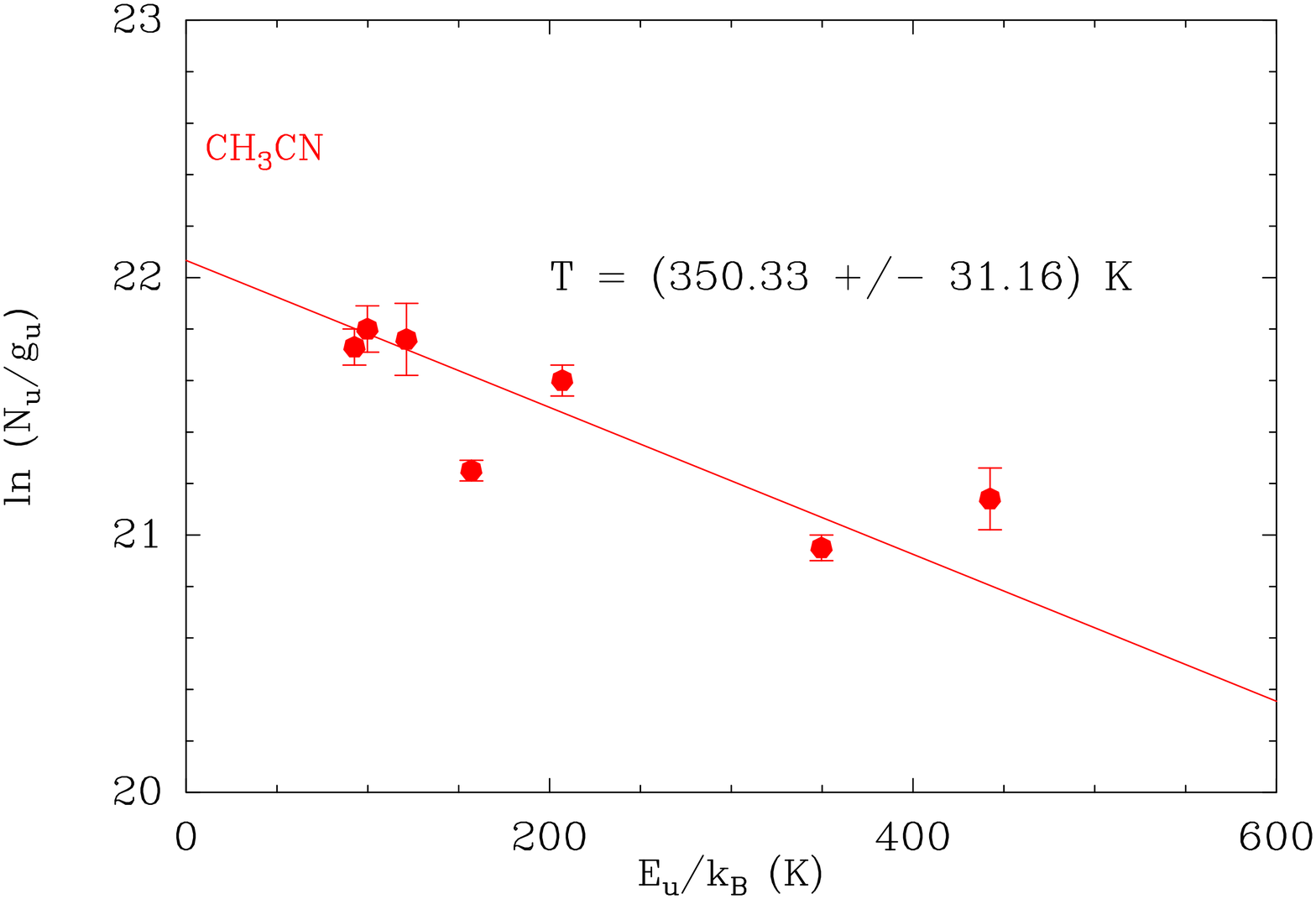}
\caption{CH$_3$CN temperature diagram in direction of NGC7538--IRS1. Red dots indicate detected and partially detected lines. Error bars (3$\sigma$) only reflect the uncertainties in the Gaussian fit of the lines. }
\label{fig:NvsQ_CH3CN}
\end{figure}

\begin{figure}
\centering
    \includegraphics[width=\columnwidth]{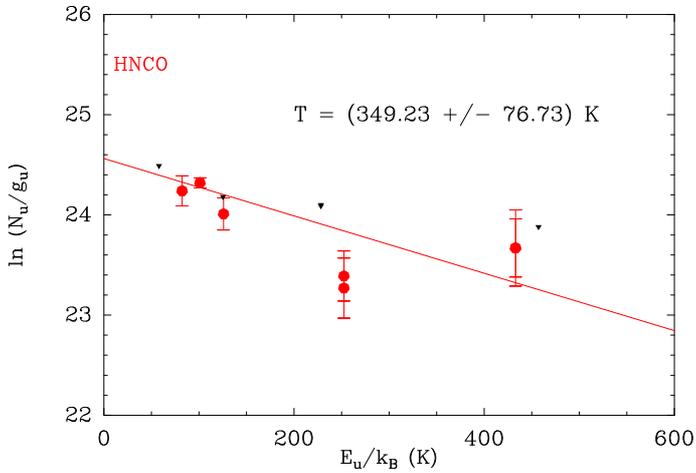}
\caption{HNCO temperature diagram in direction of NGC7538--IRS1. Red dots indicate detected and partially detected lines, while dark triangles mark blended lines. Error bars (3$\sigma$) only reflect the uncertainties in the Gaussian fit of the lines. }
\label{fig:NvsQ_HNCO}
\end{figure}

\begin{figure}
\centering
    \includegraphics[width=7cm,angle=270]{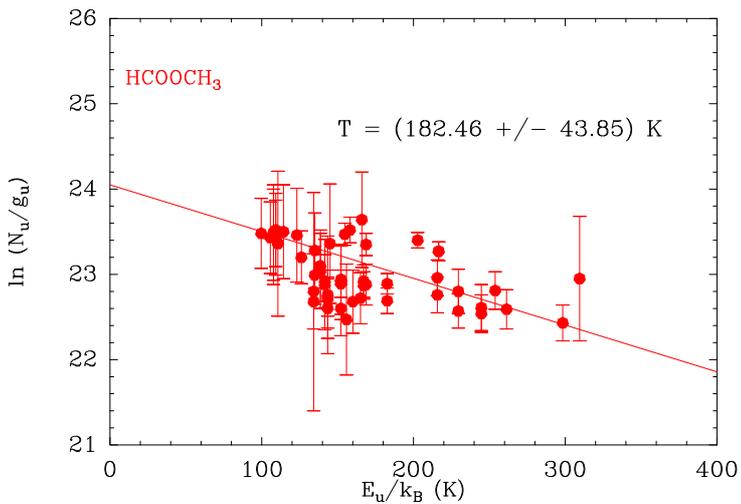}
\caption{ HCOOCH$_3$ temperature diagram in direction of NGC7538--IRS1. Red dots indicate detected and partially detected lines. Error bars (3$\sigma$) only reflect the uncertainties in the Gaussian fit of the lines. }
\label{fig:NvsQ_HCOOCH3}
\end{figure}

In our work, the rovibrational partition function is computed as the
product of the rotational contribution, given as a direct sum, the
harmonic approximation of the vibrational partition function, and, in
the case of
methyl formate, the torsional contribution computed as
  a direct sum (see above sections). To estimate the total
    column density in the survey NGC-7538, the 
  partition function in the present study has been calculated at the specific
temperatures of 350.33~K for CH$_3$CN, 349.23~K for HNCO, and 182.46~K for
HCOOCH$_3$ { (see Tables D.1, D.2 and D.3, available at the CDS),} obtained from
Figs.~\ref{fig:NvsQ_CH3CN}, \ref{fig:NvsQ_HNCO}, and
\ref{fig:NvsQ_HCOOCH3}, respectively. Moreover, because in general the
partition functions 
are reported at typically 150~K, 225~K, 300~K, and 500~K in
Table~\ref{tab-ColumnDens} we also show the results obtained after
interpolating (linear or nonlinear) their values at the excited
temperatures of the molecular species. The nonlinear interpolation of
the partition function in the present study provides  almost the
same result as the one computed.

For isocyanic acid, HNCO, we also indicate in
  Table~\ref{tab-PartF} the value of the partition function from the
  CDMS catalog~\citep{endres2016}, which provides the rotational 
contribution for the partition function; the vibrational
contribution is unfortunately missing. 
 For methyl cyanide, CH$_3$CN, the CDMS catalog provides the rovibrational
contribution for the partition function but only the lowest
frequencies of the vibrational fundamental modes up to 1200~cm$^{-1}$
are taken into account. Finally, for methyl formate, HCOOCH$_3$,
the JPL catalog provides a value for the rotational  and
torsional contributions of the partition function but only considering
the two lowest torsional states v$_{\rm t}=0$ and $1$.

\begin{figure*}
\centering
    \includegraphics[width=16cm]{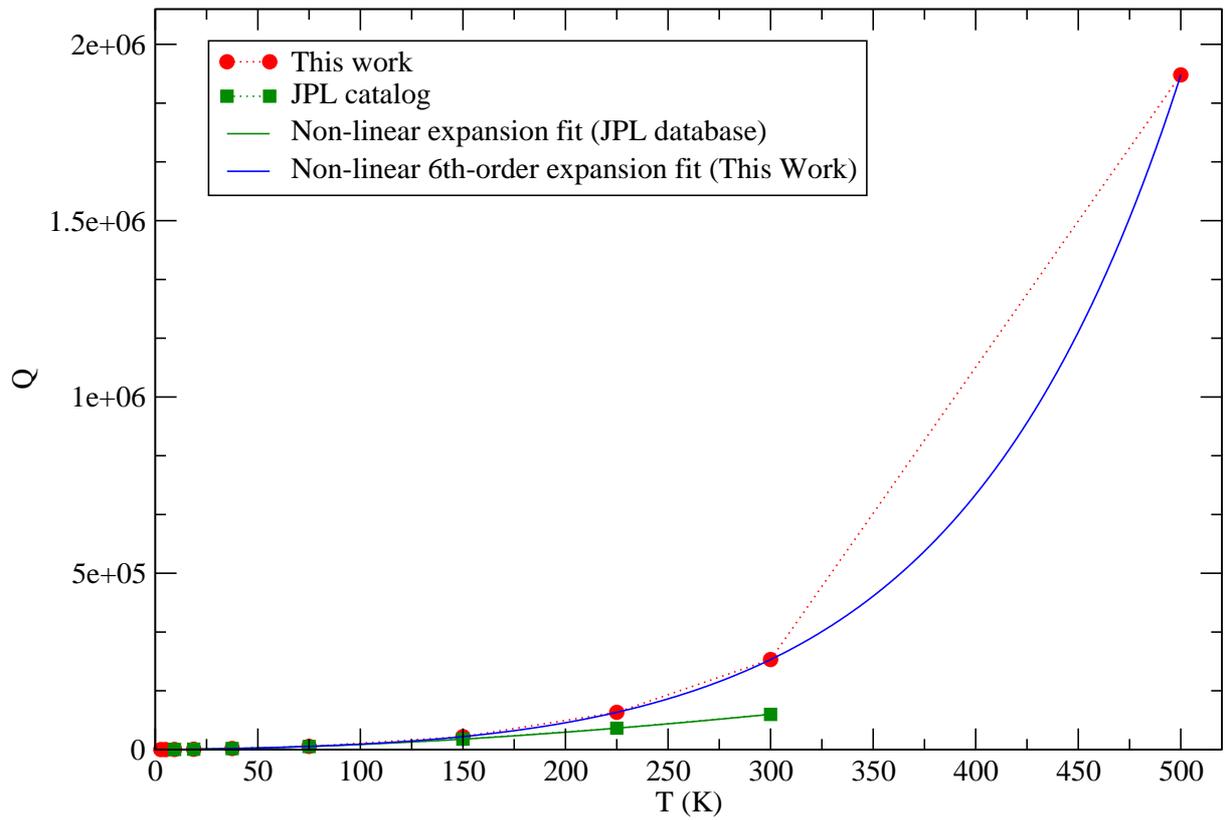}
\caption{Convergence study of the partition function for methyl
  formate. Comparison of the partition function values computed
  in the present study (full red circles) and JPL database (full
  green squares). The linear interpolation of the partition function
  computed in the present study is
  given as red dashed lines while the nonlinear sixth-order polynomial
  fit is given as a blue line. From JPL, the linear interpolation and
  the nonlinear fit (green line) are apparently overlapped.
The differences between the two partition functions are due to
  the fact that the JPL partition function does not include the
  vibrational contribution and does take into account the torsional
  contribution from Eq.~(\ref{Qtor}) only up to v$_{\rm t}= 0$ and $1$, whereas we 
are including torsional levels up to v$_{\rm t}= 10$ as well as the
vibrational contribution.
}
\label{fig:Q-MF}
\end{figure*}

It is notable that the difference between our calculated partition
function values and that from the databases are rather large, especially for
high temperatures (i.e., $\ge$ 200-300~K; see Table~\ref{tab-PartF} for
  CH$_3$CN and HNCO and Fig.~\ref{fig:Q-MF} for HCOOCH$_3$). The
resulting differences for the derived 
ISM molecular column densities are of the same order, from 9\%
up to about 40\%, as shown in Table~\ref{tab-ColumnDens}. 
In addition, a salient point of Table~\ref{tab-ColumnDens} is that
the relative differences between the column densities estimated using
our partition function and that of the CDMS/JPL database  are less
important in the case of a  nonlinear interpolation approximation than in the case of a linear one (9\%, 21\%, and 35\% vs. 18\%,
  30\%, and 43\% for CH$_3$CN, HNCO, and HCOOCH$_3$, respectively),
reflecting the importance of the contribution of the vibrational
modes. This trend is observed for all three targeted O-bearing and
N-bearing molecules (see Table~\ref{tab-ColumnDens}). Incidentally, it
is important to note that for molecules with low  vibrational energy modes and
large amplitude motions (large amplitude motions give rise to low-lying torsional energy modes) the effect of using a nonconvergent
partition function is also quite drastic. 

The importance of undertaking a convergence study of
the partition function in the characteristic temperature range of ISM
was highlighted for the first time in 2014 when methyl formate (HCOOCH$_3$)
was identified in Orion-KL from the ALMA Science Verification
Observations~\citep{favre2014}. In that study,
the methyl formate column density was derived and the values were higher by a
factor of between two and five with respect to previous reported
values~\citep{favre2011}. Such a discrepancy is almost of the same order of
magnitude as the observational uncertainties resulting from the
instruments, the data calibration, and analysis. 
Nonetheless, the ISM molecular abundances are used to constrain the
astrochemical models, which aim to investigate their formation
pathways. It is therefore crucial to involve a converged partition
function and to test the adequacy of the various approximations
  made in the partition function to derive more accurate abundances,
especially in the case of isotopolog species. Isotopic ratios,
  obtained computing the relative abundance between two isotopic
  molecular species, are
valuable for understanding the chemical evolution of interstellar
material together with the impact of interstellar gas-phase chemistry
processes that may occur during their formation \citep[e.g.,
see][]{Charnley:2004,Wirstrom:2011,favre2014}. Therefore, the
approximation order of the partition function and its level of
  convergence carried out to estimate isotopic ISM abundance ratios is relatively important and must be
consistent for all the isotopologs. An incorrect partition function,
or the use of different  approximations for the partition
functions between the isotopologs of a same molecular species, will
result in incorrect isotopic ratios 
and therefore erroneous astronomical conclusions \citep[e.g.,
see][]{favre2014}. A comprehensive convergence study of the molecular
partition function in the temperature range of ISM is therefore a
necessity. 

Finally, we emphasize that the vibrational partition function
  contributions used in our present study (see
Tables~\ref{tab-PartF} and \ref{tab-PartF-MF}) are still incomplete
because we use the harmonic approximation even though
  this could represent a good approximation, as was mentioned in
  Sect.~\ref{subsec-Qvib} when considering H$_2$S as an example.
This deficiency could be overcome either by implementing the anharmonicity
of the vibrational modes in the equation for the vibrational
  partition function or by using its direct sum
expression for the potentially populated vibrational energy term values in the
ISM temperatures. 
At the present time, there is unfortunately no vibrational partition function
analytical equation involving the anharmonicities, nor are there often complete vibrational analyses for the most relevant
astrophysical molecules, even though these would significantly
benefit the partition function calculation. Indeed, the anharmonicity usually
decreases the vibrational term values with respect to the harmonic
values by increasing the exponential terms of the direct sum of
the vibrational partition functions, in comparison with the harmonic
approximation. As a consequence, the molecular column densities could
generally be even slightly higher if the anharmonicity were taken into
account~\citep{skouteris2016}.\\

\section{Conclusions}
\label{sec-conclusions}

In the present study we investigate the impact of different
approximations of the partition function together with some interpolation
and/or extrapolation procedures on  estimations of the interstellar molecular column
density.  For that purpose, we used astronomical
observations of N- and O-bearing molecules with different
symmetries. Our analysis shows that different methods can lead to a
relative difference of up to 43\%.

Our analysis has shown that considerable errors
  in the determination of the molecular column density can be overcome
  considering the following items.

\begin{enumerate}
\item An appropriate convergence study of the partition function should
  be carried out in the temperature range of interest for the
  astronomical system under study in order to test the quality of
    the various approximations made while computing the partition function.

\item The value of $g_{\rm ns}$ used in the calculation of $Q$
  should be checked before substituting it in the intensity equation.
\item It could be beneficial to use precise interpolation and
  extrapolation approaches, that is, nonlinear fitting, for providing a
  rotational-vibrational(-torsional) partition function as a function of the temperature when the
  molecular column density is determined from the temperature
  diagram. Indeed, as shown in our study, a linear interpolation overestimates the partition function which leads to overestimation of the resulting
    molecular column densities. 
\end{enumerate}

It is important to emphasize that for computing a complete 
  partition function, comprehensive
  information of the rotational, vibrational, and torsional  (if
  appropriate) energy
  levels according to the excited temperature
  of the survey is necessary. In addition, the more complete the data set, the better the
  accuracy obtained for the partition function. Nevertheless,
  because of the lack of available spectroscopic data in the literature, one could take advantage of the present study to
    complete the partition functions of the molecules which could
      be  incompletely converged for high-temperature ISM
sources. It should be noted that in spite of the unfavorable
  estimate of the partition function 
  uncertainties considering large uncertainties for the
  rotational-vibrational-torsional data, their relative values for the
  targeted molecules are small concerning the considered convergence
  limit smaller than 1\%.

An extension of our work to other molecular species and surveys will 
 be carried out in the near future, for example for propyne
 (CH$_3$CCH), for which the present partition function only considers the
vibrational states up to 700~cm$^{-1}$~\citep{muller2002}, or acetone
(CH$_3$C(O)CH$_3$), and acetamide (CH$_3$C(O)NH$_2$), for which
  only the lowest torsional states are taken into account. This new
study will allow us to verify if the 
partition functions converge in the temperature range of the surveys
under study. In addition, these can be tabulated with polynomial
coefficients to obtain an appropriate interpolation or extrapolation
at the temperature of the survey. This extension will be particularly
useful for surveys of sources with temperatures higher than 300~K 
  which can be targeted in the future and in addition this could be relevant to future
  observations in the infrared, for example by { the James Webb Space
    Telescope (JWST), SOFIA,  the Space Infrared Telescope for Cosmology
  and Astrophysics (SPICA), and the Origins Space Telescope (OST).}

\begin{acknowledgements}

This work is partly supported by CMST COST Action CM1401 Our Astro-Chemical
History and CMST COST Action CM1405 MOLIM.
The work of CF is supported by the French National Research Agency in the
framework of the Investissements d'Avenir program (ANR-15-IDEX-02),
through the funding of the ``Origin of Life'' project of the
Univ. Grenoble-Alpes. CC and CF acknowledge funding from the
European Research Council (ERC) under the European Union’s Horizon
2020 research and innovation programme, for the Project "The Dawn of
Organic Chemistry" (DOC), grant agreement No 741002. 
IK would like to thank the French programme of
  Chimie Interstellaire PCMI.
CF and DF acknowledge support from the Italian Ministry of
Education, Universities and Research, project SIR (RBSI14ZRHR). EAB
acknowledges support from NSF (AST-1514670) and NASA
(NNX16AB48G). MC acknowledges the financial support from
  FIS2014-53448-C2-2-P (MINECO, Spain) and from the Consejer\'{\i}a de Conocimiento, Investigaci\'on y Universidad, Junta de Andaluc\'{\i}a and European Regional Development Fund (ERDF), ref. SOMM17/6105/UGR.

\end{acknowledgements}

\bibliographystyle{aa}
%\bibliography{AA_20190520}

%%%%%%%%%%%%%%%%% TABLES %%%%%%%%%%%%%%%%%%%%%

\begin{table}
\caption{Rotational partition function for isocyanic acid
  (HNCO). Comparison between the direct sum values and the classical
  approximation$^a$.}
\label{tab-Qrot-HNCO}
\begin{center}
{\footnotesize
\begin{tabular}{cccc}
\hline\hline
$T$(K) & $Q_{\rm rot}^{\rm approx}$$^b$ & $Q_{\rm rot}(\mbox{\rm direct sum})^c$ & Rel. Diff.(\%)$^d$ \\ \hline

2.725&     2.43   &   5.51 &55.92\\
  5.0  &   5.86   &   9.82 &40.32\\
  9.375&  14.80   &  18.45 &19.76\\
 18.75 &  41.48   &  42.83 & 3.15\\
 37.50 & 116.77   & 117.31 & 0.46\\
 75.0  & 329.50   & 331.99 & 0.75\\
150.0  & 930.86   & 943.72 & 1.36 \\
225.0  &1709.42   &1742.08 & 1.87\\
300.0  &2631.32   &2692.23 & 2.26\\
500.0  &5660.36   &5810.29 & 2.58\\
\hline\hline
\end{tabular}
\begin{flushleft}
$^a$ Nuclear spin degeneracy is considered as $1$ (for
  more details, see Appendix~\ref{appendix-nuclear-spin}).\\
$^b$ Rotational partition function computed with approximated
rotational partition function for slightly asymmetric tops
given by \citet{mcdowell1990}:

{\tiny
$Q_{rot}^{\mbox{McDowell}}(T)  \approx g_{\rm ns} \, \sqrt{{\pi \over
    A \,B \,C}\,\left({k\,T \over h}\right)^3}\,e^{\left(h \, B
  \left(4 - {B^2 \over  A\,C }\right) \over 12\,k\,T \right)} \left(1 + \frac{1}{90}\,\left({h \,B \,\left(1 -
  {B^2 \over A \,C }\right) \over k \, T}\right)^2\right)
$}

$^c$ Rotational partition function computed as a direct sum with
Eq.~(\ref{Qrot}) considering the predicted rotational energy levels up to
  J=135 and K$_a$=30~\citep{lapinov2007}.\\

$^d$ Relative differences.

\end{flushleft}
}
\end{center}
\end{table}

\begin{table*}
\caption{Vibrational and rotational-vibrational partition function for methyl cyanide
  (CH$_3$CN) and isocyanic acid (HNCO). Comparison between the values
  obtained in the present study and those published in CDMS catalog$^a$.}
\label{tab-PartF}
\begin{center}
%{\footnotesize
{\tiny
\begin{tabular}{ccccc|cccc}
\hline\hline
&\multicolumn{4}{c}{CH$_3$CN} &   \multicolumn{4}{c}{HNCO} \\ 
\hline
$T$(K) &$Q_{\rm vib}^{\rm harm}$~$^b$ & $Q_{\rm rv}$(Present
work)$^c$ &$Q$(CDMS)$^d$ & Rel. Diff.(\%)$^e$ &$Q_{\rm vib}^{\rm harm}$~$^b$ & $Q_{\rm rv}$(Present work)$^f$ &
$Q$(CDMS)$^g$ & Rel. Diff.(\%)$^e$\\ \hline
  2.725&1.000000  &     7.3186(57)   &    10.38& -41.80&1.000000  &    5.5129(50)   &   5.51 & 0.00\\
  5.0  &1.000000  &    16.8274(53)   &    21.37& -26.98&1.000000  &    9.8228(42)   &   9.82 & 0.00\\
  9.375&1.000000  &    42.4937(55)   &    48.07& -13.13&1.000000  &   18.4493(36)   &  18.45 & 0.00\\
 18.75 &1.000000  &   119.2669(59)   &   123.24& -3.33&1.000000  &    42.8295(30)   &  42.83 & 0.00\\
 37.50 &1.000002  &   336.0668(62)   &   336.81& -0.22&1.000000  &  117.3053(27)   & 117.30 & 0.01\\
 75.0  &1.001822  &   950.611(24)   &   950.75&-0.01&1.000019  &  332.0002(27)   & 331.99 & 0.00\\
150.0  &1.063424  &  2852.4(12)   &  2855.81&-0.12&1.006395  &  949.759(40)   & 943.71 &  0.64 \\
225.0  &1.233076  &  6076.5(59)   &  6033.91& 0.70&1.048753  & 1827.01(36)   &1742.43 &  4.63\\
300.0  &1.507618  & 11441(16)   & 11012.72& 3.74&1.144368  & 3080.9(12)   &2695.34 & 12.51\\
500.0  &2.970625  & 48538(107)   & 36469.42&  24.86&1.667813  & 9690.5(89)   &5866.52 & 39.46\\
\hline\hline
\end{tabular}

\begin{flushleft}
$^a$ The  nuclear spin degeneracy was considered as $1$.

$^b$ The vibrational partition function was computed with the harmonic
approximation given by Eq.~(\ref{Qvib-harm}).

$^c$ $Q_{\rm rv}=Q_{\rm rot}(\mbox{\rm direct sum}) \, Q_{\rm
    vib}^{\rm harm}$. An upward estimate of the uncertainties are
  given in parentheses in 
  units of the last quoted digits.

$^d$ Partition function computed as a product
of the direct sum of the rotational contribution and the vibrational
contribution, which considers vibrational
fundamental levels up to about
1200~cm$^{-1}$~\citep{endres2016}.

$^e$ Relative difference of the partition function given in the present study
with respect to the values in CDMS catalog.

$^f$ $Q_{\rm rv}=Q_{\rm rot}(\mbox{\rm direct sum}) \, Q_{\rm
    vib}^{\rm harm}$. An upward estimate of the uncertainties is given in parentheses in
  units of the last quoted digits.

$^g$ This is a rotational partition function computed as a
direct sum~\citep{endres2016}.

\end{flushleft}
}
\end{center}
\end{table*}

\begin{table*}
\caption{Rotational partition function for methyl cyanide
  (CH$_3$CN). Comparison between the direct sum values and the
  classical approximation$^a$.}
\label{tab-Qrot-CH3CN}
\begin{center}
{\footnotesize
\begin{tabular}{cccc}
\hline\hline
$T$(K) & $Q_{\rm rot}^{\rm approx}$~$^b$ & $Q_{\rm
  rot}(\mbox{\rm direct sum})^c$ & Rel. Diff.(\%)$^d$ \\ \hline

2.725  &    6.92  &     7.32  & 5.49\\
  5.0  &   16.78  &    16.83  & 0.30\\
  9.375&   42.49  &    42.49  & 0.01\\
 18.75 &  119.26  &   119.27  & 0.01\\
 37.50 &  336.00  &   336.07  & 0.02\\
 75.0  &  948.52  &   948.88  & 0.04\\
150.0  & 2680.23  &  2682.27  & 0.08\\
225.0  & 4922.31  &  4927.94  & 0.11\\
300.0  & 7577.17  &  7588.72  & 0.15\\
500.0  &16300.33  & 16339.16  & 0.24\\
\hline\hline
\end{tabular}

\begin{flushleft}
$^a$ Nuclear spin degeneracy is considered as $1$ (for
  more details, see Appendix~\ref{appendix-nuclear-spin}).\\

$^b$ Rotational partition function computed with approximated
rotational partition function for symmetric tops given 
by \citet{mcdowell1990}:

{\tiny
$Q_{\rm rot}^{\rm McDowell}(T)  \approx g_{\rm ns} \, \sqrt{{\pi \over
    A \,B^2}\,\left({k\,T \over h}\right)^3}\,e^{\left(h \, B
  \left(4 - {B \over  A }\right) \over 12\,k\,T \right)} \left(1 + \frac{1}{90}\,\left({h \,B \,\left(1 -
  {B \over A }\right) \over k \, T}\right)^2\right)
$}
\\

$^c$ Rotational partition function computed as a direct sum with
Eq.~(\ref{Qrot}) using all the predicted rotational energy levels up
to J=99~\citep{muller2009}.\\

$^d$ Relative differences.

\end{flushleft}
}
\end{center}
\end{table*}

\begin{table*}
\caption{Rotational partition function for methyl formate
  (HCOOCH$_3$). Comparison between the direct sum values and the
  classical approximation$^a$.}
\label{tab-Qrot-MF}
\begin{center}
{\footnotesize
\begin{tabular}{cccc}
\hline\hline
$T$(K) & $Q_{\rm rot}^{\rm approx}$~$^b$ & $Q_{\rm
    rot}(\mbox{\rm direct sum})^c$ & Rel. Diff.(\%)$^d$ \\ \hline

  2.725&    27.95  &    28.90 &  3.30\\
  5.0  &    69.47  &    70.81 &  1.90\\
  9.375&   178.35  &   180.38 &  1.12 \\
 18.75 &   504.45  &   507.91 &  0.68 \\
 37.50 &  1426.80  &  1433.55 &  0.47 \\
 75.0  &  4035.60  &  4051.35 &  0.39 \\
150.0  & 11414.40  & 11459.31 &  0.39 \\
225.0  & 20969.59  & 21052.21 &  0.39 \\
300.0  & 32284.79  & 32347.21 &  0.19 \\
500.0  & 69465.82  & 67714.62 & -2.59\\
\hline\hline
\end{tabular}

\begin{flushleft}
$^a$ Nuclear spin degeneracy is considered as $1$ (for
  more details, see  Appendix~\ref{appendix-nuclear-spin}).\\

$^b$ Rotational partition function computed with the classical approximation
for slightly asymmetric tops~\citep{Herzberg}.
For HCOOCH$_3$ we computed this rotational
partition function using the rotational parameters $A$, $B$ and $C$ of
\citet{tudorie2012} in the Principal Axis System.

$^c$ Rotational partition function computed as a direct sum with Eq.~(\ref{Qrot}). The
rotational energies (A-symmetry v$_{\rm t}=0$ levels) are computed up to
$J=79$ with the Hamiltonian parameters of \citet{carvajal2007}.\\

$^d$ Relative differences.

\end{flushleft}
}
\end{center}
\end{table*}

\begin{table*}
\caption{Torsional$^a$ and vibrational contributions of the partition
  function for methyl formate (HCOOCH$_3$).}
\label{tab-Qtor-Qvib-MF}
\begin{center}
{\footnotesize
\begin{tabular}{ccccc|c}
\hline\hline
T(K) & $Q_{\rm tor}^{4}$ $^b$ &
$Q_{\rm tor}^{6}$ $^b$ &
$Q_{\rm tor}^{8}$ $^b$ &
$Q_{\rm tor}^{10}$ $^{b,c}$ & $Q_{\rm vib}^{\rm harm}$ $^d$\\ \hline

  2.725& 1.99290 & 1.99290 & 1.99290 & 1.99290 & 1.00000 \\
  5.0  & 1.99612 & 1.99612 & 1.99612 & 1.99612 & 1.00000 \\
  9.375& 1.99793 & 1.99793 & 1.99793 & 1.99793 & 1.00000 \\
 18.75 & 1.99904 & 1.99904 & 1.99904 & 1.99904 & 1.00000 \\
 37.50 & 2.01230 & 2.01230 & 2.01230 & 2.01230 & 1.00001 \\
 75.0  & 2.18594 & 2.18594 & 2.18594 & 2.18594 & 1.00397 \\
150.0  & 2.91686 & 2.92139 & 2.92175 & 2.92178 & 1.09599 \\
225.0  & 3.73488 & 3.77670 & 3.78442 & 3.78584 & 1.32486 \\
300.0  & 4.45748 & 4.58622 & 4,62247 & 4,63267 & 1.70330 \\
500.0  & 5.84224 & 6.34526 & 6.58039 & 6.69031 & 4.11706 \\
\hline\hline
\end{tabular}

\begin{flushleft}
$^a$ Torsional partition functions computed as a
  direct sum by
considering a number of torsional energies up to a maximum quantum
number v$_{\rm t}^{\rm max}$ in Eq.~(\ref{Qtor}).

$^b$ This approximation considers the predicted torsional
energies up to v$_{\rm t}=4$ from the Hamiltonian parameters of
\citet{tudorie2012} and the harmonic estimates~\citep{favre2014} for  the
torsional energies from  v$_{\rm t}=5$ to v$_{\rm t}=10$.

$^c$ This result, with torsional energies up to v$_{\rm t}=10$, is used as a
final result in the present study.

$^d$ The vibrational partition function was computed with the harmonic
approximation given by Eq.~(\ref{Qvib-harm}).

\end{flushleft}
}
\end{center}
\end{table*}

\begin{table*}
\caption{Rotational-torsional-vibrational partition function$^a$
  for main isotopolog of methyl formate (H$^{12}$COO$^{12}$CH$_3$).} 
\label{tab-PartF-MF}
\begin{center}
{\footnotesize
\begin{tabular}{crrc}
\hline\hline        
$T$(K) & \citet{favre2014}$^b$ & Present work$^c$ & Rel. Diff. (\%)$^d$\\ \hline
  2.725&       --- &       57.60(22)  & --- \\
  5.0  &       --- &      141.35(29)  & --- \\
  9.375&    360.33 &      360.38(39)  & 0.01\\
 18.75 &   1015.31 &     1015.33(55)  & 0.00\\
 37.50 &   2885.30 &     2884.75(78)  &-0.02\\
 75.0  &   8894.06 &     8891.2(22)  &-0.03\\
150.0  &  36433.43 &    36695.5(552)  & 0.71\\
225.0  & 104015.96 &   105592(261)  & 1.49\\
300.0  & 249172.44 &   255246(788)  & 2.38\\
500.0  &       --- &  1913396(7629)  & ---\\
\hline\hline
\end{tabular}

\begin{flushleft}
$^a$ The nuclear spin degeneracy was not considered in these calculations.

$^b$ Partition function given in \citet{favre2014}. The
rotational partition function was computed as a direct sum (\ref{Qrot}), the
torsional contribution with Eq.~(\ref{Qtor}) using approximated
torsional energy term values up to v$_{\rm t}=6$ and vibrational contribution as in
Eq.~(\ref{Qvib-harm}). The partition function values
at temperatures of 2.725~K, 5.0~K and 500.0~K were not calculated.

$^c$ Partition function computed with Eq.~(\ref{rtvQapprox}). The
rotational partition function considered as a direct sum (\ref{Qrot}),
except for $T$=500~K (see main text). The torsional contribution is
calculated by means of Eq.~(\ref{Qtor}) using approximated torsional
energy term values up to v$_{\rm t}=10$ and vibrational
contribution as in Eq.~(\ref{Qvib-harm}).  An upward estimate of the uncertainties are given in parentheses in units of the last quoted digits.

$^d$ Relative difference of the partition function given in the present study
with respect to that of \citet{favre2014}.
\end{flushleft}
}
\end{center}
\end{table*}

\begin{table*}
\caption{Column density estimates of CH$_3$CN, HNCO, and HCOOCH$_3$ in NGC-7538 
survey according to the partition function computed in the present study (rovibrational partition functions are computed as
  the product of the rotational contribution, given as a direct sum, the harmonic approximation of the vibrational partition
  function, and, for methyl formate, the torsional
  partition sum) and in the CDMS and JPL databases. The 
    interpolated partition
functions at the excitation temperatures deduced from
Figs.~\ref{fig:NvsQ_CH3CN}, \ref{fig:NvsQ_HNCO}, and
\ref{fig:NvsQ_HCOOCH3} (350.33~K for CH$_3$CN, 349.23~K for HNCO and
182.46~K for HCOOCH$_3$) are compared with the value
computed in the present study.
}
\label{tab-ColumnDens}
\begin{center}
{\footnotesize
\begin{tabular}{cccc|ccc|c}
\hline\hline        
         & \multicolumn{3}{c}{Linear Interpolation} &
         \multicolumn{3}{c}{Nonlinear Interpolation}&  Present work$^b$\\
CH$_3$CN & Present work & CDMS & Rel.Diff.$^a$& Present work & CDMS &
Rel.Diff.$^a$&\\ \hline
$Q$($T$=350~K) &  20715.0 & 17376.9 & & 16770.2  & 15396.3
&& 16814.5\\
$N$ (cm$^{-2}$)  &   (79.6$\pm$5.3)$\times$10$^{13}$   &  $(66.8\pm4.5)\times10^{13}$  &  18\% &  (64.5$\pm$4.3)$\times$10$^{13}$    &  $(59.2\pm3.9)\times10^{13}$ &   9\%& (64.6$\pm$4.3)$\times$10$^{13}$  \\
\hline\hline
HNCO & Present work & CDMS & Rel.Diff.& Present work & CDMS &
Rel.Diff.&Present work\\ \hline
$Q$($T$=349~K) &    4700.25 &   3472.3 & &  4169.83  &
3392.7  && 4183.4\\
$N$ (cm$^{-2}$)  &    (21.8$\pm$1.9)$\times$10$^{13}$   &
$(16.1\pm1.4)\times10^{13}$  &   30\% &   (19.3$\pm$1.7)$\times$10$^{13}$
&  $(15.7\pm1.4)\times10^{13}$ &  21\%& (19.4$\pm$1.7)$\times$10$^{13}$\\
\hline\hline
 HCOOCH$_3$  & Present work &  JPL  &  Rel.Diff. & Present work &  JPL
 &  Rel.Diff & Present work\\
\hline
 $Q$($T$=182~K) &   66091.3 &  42769.4 & &  59265.8  &
41801.6  & & 59302.1\\
 $N$ (cm$^{-2}$)  &    $(18.4\pm5.4)\times10^{14}$   &
$(11.9\pm3.5)\times10^{14}$  &  43\% &  $(16.5\pm4.8)\times10^{14}$
&  $(11.6\pm3.4)\times10^{14}$ &  35\%& (16.5$\pm$4.8)$\times$10$^{14}$\\
\hline\hline
\end{tabular}
}
\end{center}

\begin{flushleft}
  {$^a$ Relative difference between the column density estimates
    obtained from the partition functions computed in the present study and
    those given in CDMS and JPL catalogs, using the formalism Rel.~Diff $= \frac{|a+b|}{(a-b)/2}$.}

$^b$ Partition function computed with Eq.~(\ref{rtvQapprox}) for
  the excitation temperatures deduced for NGC-7538 
survey and provided { in Tables D.1, D.2 and D.3, available at the CDS.}
\end{flushleft}

\end{table*}

%%%%%%%%%%%%%%%%%%%%%%%%%%%%%%%%%%%%%%%%%%%%%%%%%%
%%%%%%%%%%%%%%%%%%%%%%%%%%%%%%%%%%%%%%%%%%%%%%%%%%

%%%%%%%%%%%%%%%%% FIGURES %%%%%%%%%%%%%%%%%%%%%

\clearpage
%\newpage

%%%%%%%%%%%%%%%%%%%%%%%%%%%%%%%%%%%%%%%%%%%%%%%%%%

%%%%%%%%%%%%%%%%% APPENDICES %%%%%%%%%%%%%%%%%%%%%

\begin{appendix}

\section{Nuclear spin statistics}
\label{appendix-nuclear-spin}

In this appendix we discuss the different nuclear spin
statistical weights used in the literature for isocyanic acid,
methyl cyanide, and methyl formate.

\subsection{Isocyanic acid (HNCO)}

Isocyanic acid has a molecular symmetry ${\cal C}_{\rm s}(M)$ and
therefore the nuclear spin statistical weight results  in $g_{\rm ns}=6$
for  both the $A^{\prime}$ and  $A^{\prime\prime}$  states according to
\citet{BJbook}.  In fact, $g_{\rm ns}$ has the same value for both
symmetries, and therefore can be canceled out by the factor $Q/g_{\rm u}$ (see Eqs.~(\ref{logInt}--\ref{logInt2})). 
As a result, we can assume that $g_{\rm ns}$ equals $6/6=1$. 
Consequently, in Tables~\ref{tab-Qrot-HNCO} and \ref{tab-PartF}, the
$g_{\rm ns}$ value for
isocyanic acid is given equal to $1$, such as in the CDMS catalog. The
JPL partition 
function is however given with a reduced value of $g_{\rm ns}=3$~\footnote{The values
  of $g_{\rm ns}$ can be obtained by different procedures
  \citep[e.g., see,][]{BJbook,mangum2015} and therefore can have different values.}. Thus, JPL and CDMS partition
functions are different by a factor of three.

\subsection{Methyl cyanide (CH$_3$CN)}

The main isotopolog of methyl cyanide ($^{12}$CH$_3^{12}$C$^{14}$N) is
a molecular species with symmetry ${\cal C}_{\rm 3v}(M)$. The
nuclear spin statistical weight of each rovibrational state (symmetries $A_1$,
$A_2$ and $E$)

is $g_{\rm ns}=12$~\citep{BJbook}. As for HNCO 
(see Sects.~\ref{subsec-gu} and \ref{subsec-HNCO}), the nuclear spin
statistical weight in Eqs.(\ref{logInt} and \ref{logInt2}) can be canceled
out since it appears in the left and in the right side of the
equations.

Nevertheless, different normalized values
of $g_{\rm ns}$ can be computed according to the author. In
Table~\ref{tab-gns-CH3CN}, some of the nuclear spin degeneracy values
computed in the
literature~\citep{BJbook,mangum2015,rinsland2008,blake1987,turner1991}
are provided.  
The difference between all those values arises from the
normalization factor (see Table~\ref{tab-gns-CH3CN}) used in the nuclear spin weight equation
and its calculation is beyond the topic of this paper. Therefore,
it is important to know the precise values of $g_{\rm ns}$ considered to compute the partition function (Eq. \ref{rtvQ}),
because the same values of $g_{\rm ns}$ should also be substituted into
the $g_{\rm u}$ factor (Eq.~\ref{gu}) in the intensity
expression~(Eqs.~\ref{logInt} or
  \ref{logInt2}). Indeed, if the partition function of Eq.~(\ref{rtvQ}) is taken from a given
reference and the nuclear spin weights from another and their
  values are different, this could give
rise to erroneous astrochemical conclusions. 

For instance, the rotational partition
function calculated as a direct sum is $Q_{\rm rot}(\mbox{\rm direct sum})
\approx 7588.7$ at $T$=300~K assuming
$g_{\rm ns}=1$ (see Table~\ref{tab-Qrot-CH3CN}) whereas the one 
published by the JPL and CDMS catalogs is $\approx 10118$
($4/3\,Q_{\rm rot}(\mbox{\rm direct sum})$) which also differs from the one calculated by \citet{rinsland2008}:
$\approx 60709$ ($8\,Q_{\rm rot}(\mbox{\rm direct sum})$). Although the rotational partition function of 
\citet{rinsland2008} is six times greater than that given by the
JPL and CDMS databases and eight times greater than that used in the present study,
all three values for the partition function agree with each other, after observing that
  they only differ by the normalization factor, which results from the
different way of computing the nuclear spin statistical weight.

\begin{table}
\caption{Nuclear statistical weight $g_{\rm ns}$ values for all symmetries
  of methyl cyanide according to the reference}
\label{tab-gns-CH3CN}
\begin{center}
  {\footnotesize
\begin{tabular}{cc}
\hline\hline
\citet{BJbook} & 12 \\
JPL/CDMS/\citet{mangum2015}&  4/3\\
\citet{rinsland2008} & 8 \\
\citet{blake1987} & 2/3 \\
\citet{turner1991} & 1/3$^a$ \\
\hline\hline
\end{tabular}
}
\end{center}

$^a$ According to \citet{turner1991} and \citet{loren1984}, the
  values of $g_{\rm ns}=g_{\rm I} \, g_{\rm K}$ (Eq.\ref{gns}) for the rotational
  states with quantum number $K=0$ or $K\ne 3n$ ($n$ is an integer)
  are half the values of $K=3n$ states. This happens if the two $K=3n$
  rotational states (with $A_1$ and $A_2$ symmetries) are considered as degenerate K-doublet and,
  in this case, the value of $g_{\rm ns}$  of $K=3n$ states should be two
  times the value of $g_{\rm ns}$ for the rotational states split into
  $A_1$ and $A_2$ species. Therefore, {in case} the $A$ symmetry
    rotational states are considered degenerated for this molecular
    species, e.g., at low temperatures, it could be necessary to
    distinguish between the
    $K=3n$-doublet, considered as ortho, and the $K\ne 3n$ states
    (with E symmetry), considered as para~\citep{loren1984}.

\end{table}

\subsection{Methyl formate  (HCOOCH$_3$)}

The main isotopolog of methyl formate (H$^{12}$COO$^{12}$CH$_3$)
belongs to the molecular symmetry group ${\cal C}_{\rm 3v}(M)$. The
nuclear spin statistical weight $g_{\rm ns}$ for each rovibrational state (symmetries $A_1$,
$A_2$ and $E$) is equal to $8$~\citep{BJbook}. Because it has the
same value for every symmetry species states, $g_{\rm ns}$ can
be considered equal to $1$ in the intensity
(Eqs.~\ref{logInt}--\ref{logInt2}). Therefore the partition function
given in Tables~\ref{tab-Qrot-MF} and
  \ref{tab-PartF-MF} and plotted in
Fig.~\ref{fig:Q-MF} does not take into account the nuclear spin
statistical weight.

As for isocyanic acid and methyl cyanide, the
nuclear spin degeneracy depends on the definition used by the author. For example,
according to \citet{blake1987} and \citet{turner1991}, a
normalized $g_{\rm ns}$ value equal to $2$ is considered for the nondegenerate
A-type as well as for the doubly degenerate
E-type species~\footnote{\citet{blake1987}
  and \citet{turner1991} consider that  $g_{\rm K}=1$ and $g_{\rm I}=2$ for the
  nondegenerate A-type and $g_{\rm K}=2$ and $g_{\rm I}=1$ for the doubly degenerate
E-type species so, according to
  Eq.~(\ref{gns}), $g_{\rm ns}= g_{\rm I} \, g_{\rm K} =2$.}, such as
is considered in the JPL catalog.

\section{Comparison of partition functions from nonlinear
    polynomial expansions and direct sums}
\label{appendix-non-linear-polyn}

In this appendix we provide the polynomial coefficients of
Eqs.~(\ref{Qpolynomial}) and (\ref{logQpolynomial}) which are fitted to
the  rotational-vibrational(-torsional) partition function values
computed in the present study for isocyanic acid (HNCO),  methyl
cyanide (CH$_3$CN), and methyl formate (H$^{12}$COO$^{12}$CH$_3$).

\subsection{Isocyanic acid}

The rovibrational partition function values of the
  present study given in Table~\ref{tab-PartF}
have been fitted to the polynomial expansions given in
Eqs.~(\ref{Qpolynomial}) and (\ref{logQpolynomial}). The parameter
values of the two nonlinear fits are given in
Table~\ref{tab-PartF-polynomial-HNCO}.  To check the
agreement between the rovibrational partition function computed 
  using direct sums with the values provided by the polynomial
expansions, their relative
differences are also provided in
Table~\ref{tab-PartF-polynomial-HNCO}. From
  Table~\ref{tab-PartF-polynomial-HNCO} we can see that the
sixth-order logarithm polynomial $\log_{10} Q(T)$ of 
Eq.~(\ref{logQpolynomial}) has better agreement in general with the
direct sum results than the $Q(T)$ expansion of
Eq.~(\ref{Qpolynomial}), in particular at low temperatures from 2.725~
to 37.5~K. Above 37.5~K the logarithm expansion remains in reasonable
agreement with $Q_{\rm rv}$, although the polynomial expansion of
$Q(T)$ is better.

Therefore, according to the small relative differences,
both polynomial expansions can be considered to interpolate the
rovibrational partition function of isocyanic acid at any temperature
from 2.725~ to 500~K. Hence, according to the better agreement of the
fitted polynomial expansion at higher temperatures 
 with respect to the nonlinear fit of the logarithm polynomial
 expansion, we decided to use the former one to
 interpolate the partition function at the temperature of  NGC-7538
 survey and so estimating the molecular column
   density (see
  Sections~\ref{sec-astro} and~\ref{sec-discussion}).

\subsection{Methyl cyanide}

As in Sect.~\ref{subsec-HNCO}, the rovibrational partition function
values of the present study for CH$_3$CN (see Table~\ref{tab-PartF}) 
have been fitted to the polynomial expansions given in
Eqs.~(\ref{Qpolynomial}) and (\ref{logQpolynomial}). The parameter
values of both nonlinear fits and the relative differences of the
fitted polynomial expansions with respect to the rovibrational
partition function values computed in the present study are given in
Table~\ref{tab-PartF-polynomial-CH3CN}. On the one hand, the
  sixth-order expansion of Eq.~(\ref{Qpolynomial}) presents
  a good agreement except for temperatures below 10~K. On the other
  hand, the logarithm polynomial of Eq.~(\ref{logQpolynomial}) in
  general give a
  reasonable agreement with $Q_{\rm rv}(\mbox{\rm Present
    work})$. Therefore, the two expansions can be used to interpolate the
rovibrational partition function of methyl cyanide at any temperature
from 10~K to 500~K. 

In the present study, we have used the logarithm
  polynomial expansion to interpolate the rovibrational partition
  function of methyl cyanide at the 
  temperature of  NGC-7538 survey  (see
  Sections~\ref{sec-astro} and~\ref{sec-discussion}) because it is better in
  general.

\subsection{Methyl formate}

The rotational-vibrational-torsional
partition function 
values of methyl formate computed in this study (see column 3 of
Table~\ref{tab-PartF-MF}), has been fitted with the two nonlinear polynomial
expansions (see Eqs.~(\ref{Qpolynomial}) and
(\ref{logQpolynomial})). The parameter values of these two fits are
given in Table~\ref{tab-PartF-polynomial-MF}. Table~\ref{tab-PartF-polynomial-MF}
also shows the
relative differences between the values of the
rotational-vibrational-torsional partition function computed in the
present study using the direct
sum and the values of the partition function fitted using polynomial values. This comparison aims to select which fitting
formula is more appropriate depending on the temperature ranges.
In the present study we have considered the two expansions, the polynomial (Eq. \ref{Qpolynomial}) and the
decimal logarithm polynomial (Eq. \ref{logQpolynomial}), up
to sixth-order. As shown in Table~\ref{tab-PartF-polynomial-MF}, 
the sixth-order polynomial from Eq.~(\ref{Qpolynomial}) undergoes the highest
deviations: 36.45\%, $-$1.88\% and $-$6.11\% at the lowest
  temperatures, $T$=2.725, 5 and 9.375~K, respectively. On the contrary, the fit of a
power expansion of $\log_{10} Q$ from Eq.~(\ref{logQpolynomial})
undergoes deviations smaller than 0.8\% for all
the temperature range (i.e. from $T$=2.725~K up to 500~K). 
However, for temperatures higher than 75~K, the sixth-order
polynomial from Eq.~(\ref{Qpolynomial}) provides deviations of the rotational-vibrational-torsional partition
function values which are smaller than
the deviations presented  by the  logarithm power expansion of
Eq.~(\ref{logQpolynomial}). Hence, we can conclude that, for
methyl formate, the temperature
polynomial expansion is more suitable to interpolate the
rotational-vibrational-torsional partition function at any temperature
above 75~K
while the decimal logarithm  $\log_{10}(T)$ polynomial
expansion seems more appropriate for interpolating the partition
function from 75~K
down to temperatures around 3~K.

\begin{table*}
\caption{Polynomial coefficients fitted to the rovibrational partition function
  values computed in the present study for isocyanic acid (HNCO)}
  \label{tab-PartF-polynomial-HNCO}
\begin{center}
{\footnotesize
\begin{tabular}{cc|cc}
\hline\hline
\multicolumn{2}{c}{Sixth-order polynomial$^a$} &
\multicolumn{2}{c}{$\log_{10}(Q)$ expansion$^b$} \\ 
\hline
$A_0$&  2.16062 &$a_0$& -0.0808404\\
$A_1$&  1.16832 &$a_1$&  3.2397\\
$A_2$&  0.0605058 &$a_2$& -4.62856 \\
$A_3$&  -0.000308594&$a_3$&  4.19036\\
$A_4$&  1.15537 $10^{-6}$&$a_4$&  -1.71058\\
$A_5$&  -1.94798 $10^{-9}$&$a_5$&  0.298682\\
$A_6$&  1.35779 $10^{-12}$ &$a_6$&  -0.0138463\\
\hline\hline
$T$ &  Rel.Diff.(\%)$^c$       & & Rel.Diff.(\%)$^c$\\
\hline
2.725 &   -5.03      & &  -0.01\\
5.0 &     3.49    & &  -0.24\\
9.375 &   1.43      & &  0.50\\
18.75 &   -1.43       & &  -0.62\\
37.50 &   0.32      & &  0.41\\
75 &     -0.04    & &  -0.06\\
150 &     0.00    & &  -0.02\\
225 &     0.00    & &  -0.11\\
300 &     0.00    & &  0.19\\
500 &     0.00    & & 0.01\\
\hline\hline
\end{tabular}

\begin{flushleft}
$^a$ Fitted coefficients of a sixth-order polynomial of  $Q$ in terms
of the temperature $T$ according to Eq.(\ref{Qpolynomial}).

$^b$ A nonlinear fit of a power expansion of
$\log_{10}(Q)$ in terms of the $\log_{10}(T)$ according to Eq.(\ref{logQpolynomial}).

$^c$ Relative difference between the rovibrational partition function
computed in the present study (Table~\ref{tab-PartF}, column 7) and the values provided
by the polynomial expansions.
\end{flushleft}
}
\end{center}
\end{table*}

\begin{table*}
\caption{Polynomial coefficients fitted to the rovibrational partition function
  values of the present study for methyl cyanide (CH$_3$CN)}
\label{tab-PartF-polynomial-CH3CN}
\begin{center}
{\footnotesize
\begin{tabular}{cc|cc}
\hline\hline
\multicolumn{2}{c}{Sixth-order polynomial$^a$} &
\multicolumn{2}{c}{$\log_{10}(Q)$ expansion$^b$} \\ 
\hline
$A_0$& -6.05528  &$a_0$& 0.412384\\
$A_1$& 3.82747  &$a_1$&  0.653017\\
$A_2$& 0.172057  &$a_2$&  1.25394\\
$A_3$& -0.00103156 &$a_3$&  -1.06785\\
$A_4$&  5.05905 $10^{-6}$&$a_4$&  0.580728\\
$A_5$&  -1.00259 $10^{-8}$&$a_5$&  -0.192079 \\
$A_6$&   8.29957 $10^{-12}$ &$a_6$&  0.0286266\\
\hline\hline
$T$ &  Rel.Diff.(\%)$^c$       & & Rel.Diff.(\%)$^c$\\
\hline
2.725 &   23.1      & &  0.07\\
5.0 &     -2.56    & &  -0.16\\
9.375 &   -3.87      & &  0.31\\
18.75 &   -0.62       & &  -0.24\\
37.50 &   0.52      & &  -0.09\\
75 &      -0.08     & &   0.37\\
150 &     0.01    & &  -0.32\\
225 &     0.00    & &  -0.07\\
300 &     0.00    & &  0.21\\
500 &     0.00    & &  -0.06\\
\hline\hline
\end{tabular}

\begin{flushleft}
$^a$ Fitted coefficients of a sixth-order polynomial of  $Q$ in terms
of the temperature $T$ according to Eq.(\ref{Qpolynomial}).

$^b$ A nonlinear fit of a power expansion of
$\log_{10}(Q)$ in terms of the $\log_{10}(T)$ according to Eq.(\ref{logQpolynomial}).

$^c$ Relative difference between the rovibrational partition function
computed in the present study (Table~\ref{tab-PartF}, column 3) and the values provided
by the polynomial expansions.
\end{flushleft}
}
\end{center}
\end{table*}

\begin{table*}
\caption{Polynomial coefficients fitted to the 
    rotational-vibrational-torsional partition function
  values of the present study for methyl formate  (H$^{12}$COO$^{12}$CH$_3$)}
  \label{tab-PartF-polynomial-MF}
\begin{center}
{\footnotesize
\begin{tabular}{cc|cc}
\hline\hline
\multicolumn{2}{c}{Sixth-order polynomial$^a$} &
\multicolumn{2}{c}{$\log_{10}(Q)$ expansion$^b$} \\ 
\hline
$A_0$&  -76.7135 &$a_0$&  1.13999\\
$A_1$&  38.4603 &$a_1$&  1.36329\\
$A_2$&  1.15805 &$a_2$&  0.111789\\
$A_3$&  -0.00445196 &$a_3$&  0.0929227\\
$A_4$&  5.47575 $10^{-5}$&$a_4$&  -0.169274\\
$A_5$&  -1.28533 $10^{-7}$&$a_5$&  0.0658082\\
$A_6$&  1.76354 $10^{-10}$&$a_6$&  -0.00419615\\
\hline\hline
$T$ &  Rel.Diff.(\%)$^c$       & & Rel.Diff.(\%)$^c$\\
\hline
2.725 &   35.96       & &  -0.02\\
5.0 &    -1.88    & &  0.27\\
9.375 &  -6.11      & &  -0.34\\
18.75 &  -1.31       & &  0.23\\
37.50 &   0.91       & &  0.54\\
75 &     -0.13     & &   -0.67\\
150 &     0.01    & &  0.41\\
225 &     0.00    & &  0.85\\
300 &     0.00    & &  -0.61\\
500 &     0.00    & &  0.28\\
\hline\hline
\end{tabular}

\begin{flushleft}
$^a$ Fitted coefficients of a sixth-order polynomial of  $Q$ in terms
of the temperature $T$ according to Eq.(\ref{Qpolynomial}).

$^b$ A nonlinear fit of a power expansion of
$\log_{10}(Q)$ in terms of the $\log_{10}(T)$ according to Eq.(\ref{logQpolynomial}).

$^c$ Relative difference between the rovibrational partition function
computed in the present study (Table~\ref{tab-PartF-MF}, column 3) and the values provided
by the polynomial expansions.
\end{flushleft}
}
\end{center}
\end{table*}

\section{Spectroscopic and observational line parameters}
\label{appendix-observational}

The spectroscopic and observational line parameters for the molecular
species HNCO, CH$_3$CN and HCOOCH$_3$ are given here.

%++++++++++++++++
% TABLE HNCO
%++++++++++++++++
\begin{sidewaystable}
\caption{Spectroscopic and Observational Line Parameters for HNCO.}
\label{tabhnco}
%\centering
\begin{tabular}{llcccccccc} 
\hline\hline             
Frequency & Transition& $E_{\rm u}$  & $A$ &  $v_{\rm LSR}$&  $\Delta v_{\rm LSR}$ &   $T_{\rm B}$&  $W$  &   $N_{\rm u}/g_{\rm u}$  & Comments\\
(MHz)  &  &  (K)  &  ($\times$ 10$^{-4}$~s$^{-1}$)  &  (km~s$^{-1}$)  & (km~s$^{-1}$) & (K) &  (K~km~s$^{-1}$)  & ($\times$ 10$^{9}$~cm$^{-2}$) & \\
(1) & (2) & (3)&(4)&(5)&(6)&(7)&(8)&(9) & (10) \\
\hline
218981.009      &10$_{1,10}$- 9$_{1, 9}$        &101.1  &       1.42    &       -59.58  (0.04)  &       4.13    (0.07)  &               0.26    (0.01)  &       1.16    (0.02)  &               3.64    (0.02)  &        PB      \\
219656.769      &       10$_{3, 8}$- 9$_{3, 7}$&433.0   &       1.20    &       -59.53  (0.25)  &       3.99    (0.50)  &               0.13    (0.00)  &       0.51    (0.06)  &               1.90    (0.13)  &       D       \\
219656.771      &       10$_{3, 7}$- 9$_{3, 6}$&433.0   &       1.20    &       -59.54  (0.20)  &       4.00 (0.41)  &               0.12 (0.02)     &       0.51    (0.05   )       &               1.91    (0.10)          &       D       \\
219733.850      &10$_{2, 9}$- 9$_{2, 8}$        &228.3  &       1.35    &               -                       &               -                       &       $\le$   0.21                            &               -                       &       $\le$   2.93                            &       B       \\
219737.193      & 10$_{2, 8}$- 9$_{2, 7}$&      228.3   &       1.35    &               -                       &               -                       &       $\le$   0.21                            &               -                       &       $\le$   2.91                            &       B       \\
219798.274      &       10$_{0,10}$- 9$_{0, 9}$&58.0    &       1.47    &                       -               &       -               &       $\le$   0.34                            &                                       -&      $\le$   4.32                            &       B       \\
262769.477      &       12$_{ 1,12}$-11$_{ 1,11}$ &125.3        &       2.48    &               -                       &-                                      &       $\le$   0.35                            &               -                       &       $\le$   3.13                            &       B       \\
263580.924      &       12$_{ 3,10}$-11$_{ 3, 9}$ &457.2        &       2.16    &               -                       &-                                      &       $\le$   0.23                            &               -                       &       $\le$   2.35                            &       B       \\
263672.912      &       12$_{ 2,11}$-11$_{ 2,10}$&252.5 &       2.37    &       -58.96  (0.18)  &       3.88    (0.48)  &               0.13    (0.02)  &       0.56    (0.06)  &               1.27    (0.10)  &       D       \\
263678.709      &       12$_{ 2,10}$-11$_{ 2, 9}$ &252.5        &       2.37    &       -59.52  (0.15)  &       3.63    (0.32)  &               0.16    (0.02)  &       0.63    (0.05)  &               1.44    (0.08)  &       D       \\
263748.625      &       12$_{ 0,12}$-11$_{ 0,11}$&82.3  &       2.56    &       -59.13  (0.11)  &       4.38    (0.25)  &               0.34    (0.03)  &       1.60    (0.08)  &               3.38    (0.05)  &       PB      \\
264693.655      &       12$_{ 1,11}$-11$_{ 1,10}$&125.9 &       2.54    &       -59.68  (0.89)  &       4,63    (0.89)  &               0.25    (0.01)  &       1.25    (0.07)  &               2.68    (0.05)  &       PB      \\
\hline \hline
\end{tabular}
\begin{flushleft}
Notes: (1) Frequencies. (2) Transition. (3) Energy of the upper level. (4) Einstein
spontaneous emission coefficient. (5)-(8) Velocity, line width at half
intensity, brightness temperature, and integrated intensities. The one
sigma uncertainties are given in brackets. (9) Column densities of the
upper state level of the transition with respect to the upper state
degeneracy $g_{\rm u}$. (10) D: detected lines, B: blended lines and PB:
partially blended lines.

\end{flushleft}
\end{sidewaystable}

%++++++++++++++++
% TABLE CH3CN
%++++++++++++++++
\begin{sidewaystable}
\caption{Spectroscopic and Observational Line Parameters for CH$_3$CN.}
\label{tabch3cn}
%\centering
\begin{tabular}{lccccccccc} 
\hline\hline             
Frequency & Transition & $E_{\rm u}$  &  $A$ &  $v_{\rm LSR}$ &   $\Delta v_{\rm LSR}$ &   $T_{\rm B}$& $W$  & $N_{\rm u}/g_{\rm u}$  & Comments\\
(MHz)  &  &  (K)  &  ($\times$ 10$^{-4}$~s$^{-1}$)  &  (km~s$^{-1}$)  & (km~s$^{-1}$) & (K) &  (K~km~s$^{-1}$)  & ($\times$ 10$^{9}$~cm$^{-2}$) & \\
(1) & (2) & (3)&(4)&(5)&(6)&(7)&(8)&(9) & (10) \\
\hline
257284.935      & 14$_{7}$-13$_{7}$     &442.4  &       1.1     &       -59.03  (0.08)  &       3.9     (0.2)   &       0.18    (0.01)  &       0.75    (0.03)          &       1.51    (0.04)  &       D       \\
257349.179      &14$_{6}$-13$_{6}$      &349.7  &       1.2     &       -59.32  (0.03)  &       3.7     (0.1)   &       0.35    (0.01)  &       1.35    (0.02)          &       1.25    (0.02)  &       D       \\
257403.584      &       14$_{5}$-13$_{5}$&271.2 &       1.29    &       -                       &        -                               &       $\le$ 0.61      &                       -               &       $\le$ 61.97                           &B      \\
257448.128      &14$_{4}$-13$_{4}$      &207.0  &       1.35    &       -59.08  (0.03)  &       3.6     (0.1)   &       0.39    (0.01)  &       1.47    (0.03)          &       2.41    (0.02)  &       D       \\
257482.791      &14$_{3}$-13$_{3}$      &157.0  &       1.41    &       -59.20  (0.02)  &       3.9     (0.1)   &       0.52    (0.01)  &       2.14    (0.03)          &       1.69    (0.01)  &       D       \\
257507.561      &14$_{2}$-13$_{2}$      &121.3  &       1.45    &       -58.94  (0.09)  &       3.7     (0.2)   &       0.47    (0.03)  &       1.84    (0.09)          &       2.83    (0.05)  &       D       \\
257522.427      &       14$_{1}$-13$_{1}$ &99.8 &       1.47    &       -58.74  (0.05)  &       3.5     (0.1)   &       0.53    (0.02)  &       1.93    (0.06)  &               2.92    (0.03)  &       PB      \\
257527.383      &14$_{0}$-13$_{0}$      &92.7   &       1.48    &       -58.89  (0.04)  &       3.6     (0.1)   &       0.48    (0.02)  &       1.83    (0.05)  &               2.74    (0.02)  &       PB      \\
\hline \hline
\end{tabular}
\begin{flushleft}
Notes: (1) Frequencies. (2) Transition. (3) Energy of the upper level. (4) Einstein
spontaneous emission coefficient. (5)-(8) Velocity, line width at half
intensity, brightness temperature, and integrated intensities. The one
sigma uncertainties are given in brackets. (9) Column densities of the
upper state level of the transition with respect to the upper state
 degeneracy $g_{\rm u}$. (10) D: detected lines, B: blended lines and PB:
partially blended lines. 
\end{flushleft}
\end{sidewaystable}

\newpage

%++++++++++++++++
% TABLE HCOOCH3
%++++++++++++++++
\begin{table*}
\caption{Spectroscopic and observational line parameters for HCOOCH$_3$.}
\label{tabhcooch3}
%\centering
\begin{tabular}{llcccccccc} 
\hline\hline             
 Frequency & Transition & $E_{\rm u}$  &  $A$ &  $v_{\rm LSR}$ &   $\Delta v_{\rm LSR}$ &   $T_{\rm B}$&  $W$  &   $N_{\rm u}/g_{\rm u}$  & Comments\\
(MHz)  &  &  (K)  &  ($\times$ 10$^{-4}$~s$^{-1}$)  &  (km~s$^{-1}$)  & (km~s$^{-1}$) & (K) &  (K~km~s$^{-1}$)  & ($\times$ 10$^{9}$~cm$^{-2}$) & \\
(1) & (2) & (3)&(4)&(5)&(6)&(7)&(8)&(9) & (10) \\
\hline
214631.686      &       17$_{5,12}$-16$_{5,11}$ E               &       107.8   &       1.36    &       -58.94  (0.29)  &       3.28    (0.75)  &       0.24    (0.05)  &       0.84    (0.16)  &       15.89   (0.19)  & D \\
214652.590      &       17$_{5,12}$-16$_{5,11}$ A               &       107.8   &       1.36    &       -58.88  (0.28)  &       3.09    (0.68)  &       0.24    (0.05)  &       0.80    (0.15)  &       15.08   (0.19)  & D \\
216210.856      &       19$_{3,17}$-18$_{3,16}$ E               &       109.3   &       1.49    &       -58.87  (0.23)  &       3.56    (0.67)  &       0.28    (0.05)  &       1.05    (0.15)  &       16.38   (0.14)  & D \\
216216.456      &       19$_{1,18}$-18$_{1,17}$ A               &       109.3   &       1.49    &       -58.96  (0.23)  &       3.24    (0.56)  &       0.28    (0.05)  &       0.96    (0.14)  &       15.04   (0.15)  & D \\
216838.846      &       18$_{2,16}$-17$_{2,15}$ A               &       105.7   &       1.48    &       -58.85  (0.22)  &       3.31    (0.56)  &       0.25    (0.04)  &       0.89    (0.13)  &       14.93   (0.14)  & D \\
218280.835      &       17$_{1,16}$-16$_{1,15}$ E               &       99.7    &       1.51    &       -58.80  (0.23)  &       3.45    (0.58)  &       0.24    (0.04)  &       0.90    (0.12)  &       15.78   (0.14)  & D \\
229404.970      &       18$_{1,17}$-17$_{1,16}$ E               &       110.7   &       1.75    &       -58.92  (0.42)  &       3.01    (1.01)  &       0.28    (0.09)  &       0.89    (0.25)  &       14.02   (0.28)  & D \\
229539.464      &       31$_{5,27}$-31$_{3,28}$ E               &       309.5   &       0.29    &       -59.42  (0.44)  &       3.62    (1,24)  &       0.04    (0.01)  &       0.16    (0.04)  &       9.28    (0.24)  & D \\
233226.747      &       19$_{4,16}$-18$_{4,15}$ A               &       123.2   &       1.82    &       -58.81  (0.28)  &       3.15    (0.68)  &       0.31    (0.07)  &       1.04    (0.19)  &       15.47   (0.18)  & D \\
233753.921      &       18$_{2,16}$-17$_{2,15}$ E               &       114.4   &       1.84    &       -58.86  (0.29)  &       3.27    (0.71)  &       0.30    (0.07)  &       1.03    (0.19)  &       16.03   (0.18)  & D \\
234508.640      &       19$_{9,11}$-18$_{9.10}$ E               &       166.0   &       1.51    &       -59,11  (0.32)  &       3.46    (0.78)  &       0.28    (0.07)  &       1.02    (0.19)  &       18.54   (0.19)  & D \\
235046.484      &       19$_{8,12}$-18$_{8,11}$ A               &       154.8   &       1.61    &       -58.98  (0.06)  &       3.39    (0.16)  &       0.25    (0.01)  &       0.91    (0.04)  &       15.55   (0.04)  & D \\
235932.349      &       19$_{7,12}$-18$_{7,11}$ A               &       145.0   &       1.71    &       -59.03  (0.37)  &       3.32    (0.93)  &       0.24    (0.07)  &       0.86    (0,20)  &       13,94   (0.23)  & D \\
244580.304      &       20$_{4,16}$-19$_{4,15}$ E               &       135.0   &       2.10    &       -59.00  (0.28)  &       4.01    (0.72)  &       0.23    (0.05)  &       0.96    (0.14)  &       12,94   (0.15)  & D \\
244594.007      &       20$_{4,17}$-19$_{4,16}$ A               &       135.0   &       2.10    &       -58.85  (0.18)  &       3.27    (0.45)  &       0.21    (0.03)  &       0.72    (0.08)  &       9.64    (0.11)  & D \\
245772.644      &       20$_{14,7}$-19$_{14,6}$ E               &       253.9   &       1.14    &       -58.69  (0.13)  &       3.66    (0.31)  &       0.08    (0.01)  &       0.32    (0.02)  &       8.09    (0.07)  & D \\
246891.602      &       19$_{2,17}$-18$_{2,16}$ E               &       126.2   &       2.18    &       -58.75  (0.17)  &       3.30    (0.41)  &       0.24    (0.03)  &       0.86    (0.09)  &       11.95   (0.10)  & D \\
247901.651      &       22$_{2,20}$-21$_{2,19}$ E               &       143.5   &       2.26    &       -58.73  (0.35)  &       3.13    (0.93)  &       0,20    (0.06)  &       0.65    (0.15)  &       7.65    (0.23)  & D \\
247907.104      &       22$_{2,21}$-21$_{2,20}$ A               &       143.5   &       2.26    &       -58.85  (0.12)  &       3.37    (0.27)  &       0.18    (0.01)  &       0.64    (0.05)  &       7.53    (0.08)  & D \\
247922.261      &       22$_{3,20}$-21$_{3,19}$ E               &       143.5   &       2.26    &       -58.74  (0.19)  &       3.32    (0.45)  &       0.17    (0.03)  &       0.62    (0.07)  &       7.30    (0.11)  & D \\
247927.668      &       22$_{1,21}$-21$_{1,20}$ A               &       143.5   &       2.26    &       -58.83  (0.18)  &       3.04    (0.40)  &       0.17    (0.03)  &       0.56    (0.07)  &       6.53    (0.12)  & D \\
249030.973      &       20$_{5,16}$-19$_{5,15}$ E               &       141.6   &       2.18    &       -58.78  (0.15)  &       2.72    (0.34)  &       0.17    (0.02)  &       0.67    (0.07)  &       9.03    (0.10)  & D \\
249047.402      &       20$_{5,16}$-19$_{5,15}$ A               &       141.6   &       2.18    &       -58.85  (0.29)  &       3.64    (0.86)  &       0.17    (0.04)  &       0.64    (0.11)  &       8.63    (0.18)  & D \\
250246.465      &       20$_{1,19}$-19$_{1,18}$ E               &       134.3   &       2.27    &       -58.83  (0.24)  &       3.16    (0.57)  &       0.18    (0.04)  &       0.61    (0.09)  &       7.97    (0.15)  & D \\
250258.373      &       20$_{3,17}$-19$_{3,16}$ A               &       134.3   &       2.27    &       -58.86  (0.61)  &       2.92    (1.38)  &       0.18    (0.09)  &       0.54    (0.23)  &       7.06    (0.43)  & D \\
257690.326      &       22$_{3,19}$-21$_{3,18}$ E               &       152.3   &       2.50    &       -58.99  (0.22)  &       3.33    (0.55)  &       0.23    (0.04)  &       0.80    (0.11)  &       9.16    (0.14)  & D \\
257699.433      &       22$_{3,20}$-21$_{3,19}$ A               &       152.3   &       2.50    &       -58.83  (0.21)  &       3.12    (0.51)  &       0.23    (0.04)  &       0.77    (0.11)  &       8.82    (0.14)  & D \\
258081.019      &       22$_{4,19}$-21$_{4,18}$ E               &       152.2   &       2.52    &       -58.91  (0.13)  &       2.55    (0.27)  &       0.21    (0.03)  &       0.57    (0.06)  &       6.52    (0.11)  & D \\
258089.497      &       22$_{2,20}$-21$_{2,19}$ A               &       152.2   &       2.52    &       -58.91  (0.09)  &       3.05    (0.21)  &       0.23    (0.02)  &       0.76    (0.04)  &       8.69    (0.05)  & D \\
258508.124      &       23$_{1,22}$-22$_{1,21}$ A               &       155.9   &       2.56    &       -59.00  (0.27)  &       2.36    (0.52)  &       0.21    (0.06)  &       0.53    (0.12)  &       5.72    (0.22)  & D \\
258756.620      &       21$_{11,11}$-20$_{11,10}$ A             &       216.6   &       1.90    &       -58.85  (0.06)  &       3.45    (0.16)  &       0.22    (0.01)  &       0.80    (0.03)  &       12.74   (0.04)  & D \\
258756.621      &       21$_{11,10}$-20$_{11,9}$ A              &       216.6   &       1.90    &       -58.85  (0.06)  &       3.45    (0.16)  &       0.22    (0.01)  &       0.80    (0.03)  &       12.74   (0.04)  & D \\
259128.123      &       21$_{10,12}$-20$_{10,11}$ A             &       202.8   &       2.03    &       -58.93  (0.05)  &       3.50    (0.15)  &       0.26    (0.01)  &       0.97    (0.03)  &       14.48   (0.03)  & D \\
259342.015      &       24$_{1,24}$-23$_{1,23}$ E               &       158.2   &       2.63    &       -59.50  (0.09)  &       3.66    (0.22)  &       0.41    (0.03)  &       1.61    (0.08)  &       16.32   (0.05)  & D \\
259499.912      &       20$_{0,20}$-19$_{2,17}$ E               &       138.7   &       2.54    &       -59.01  (0.23)  &       3.34    (0.60)  &       0.23    (0.04)  &       0.81    (0.12)  &       10.19   (0.15)  & D \\
259521.739      &       20$_{4,16}$-19$_{4,15}$ A               &       138.7   &       2.54    &       -58.88  (0.22)  &       3.51    (0.62)  &       0.23    (0.04)  &       0.86    (0.12)  &       10.82   (0.14)  & D \\
266819.364      &       22$_{4,18}$-21$_{4,17}$ E               &       160.1   &       2.75    &       -58.68  (0.19)  &       3.11    (0.43)  &       0.19    (0.03)  &       0.63    (0.08)  &       7.04    (0.12)  & D \\
268316.643      &       23$_{3,21}$-22$_{3,20}$ A               &       165.1   &       2.83    &       -59,11  (0.15)  &       2.84    (0.34)  &       0.23    (0.03)  &       0.70    (0.07)  &       7.36    (0.10)  & D \\
269078.049      &       24$_{2,22}$-23$_{2,21}$ E               &       168.8   &       2.89    &       -58.47  (0.13)  &       3.56    (0.37)  &       0.23    (0.02)  &       0.87    (0.07)  &       8.63    (0.08)  & D \\
269084.888      &       24$_{3,22}$-23$_{3,21}$ E               &       168.8   &       2.89    &       -58.17  (0.08)  &       3.67    (0.18)  &       0.36    (0.03)  &       1.40    (0.06)  &       13.89   (0.04)  & D \\
270384.916      &       22$_{15,8}$-21$_{15,7}$ E               &       298.4   &       1.60    &       -58.92  (0.42)  &       3.16    (0.97)  &       0.08    (0.01)  &       0.28    (0.02)  &       5.52    (0.07)  & D \\
270703.117      &       22$_{13,10}$-21$_{13,9}$ E              &       261.4   &       1.95    &       -58.88  (0.11)  &       2.98    (0.25)  &       0.13    (0.01)  &       0.40    (0.03)  &       6.48    (0.08)  & D \\
270915.681      &       22$_{12,10}$-21$_{12,9}$ E              &       244.8   &       2.11    &       -58.96  (0.14)  &       3.02    (0.34)  &       0.14    (0.02)  &       0.44    (0.04)  &       6.60    (0.09)  & D \\
270939.232      &       22$_{12,11}$-21$_{12,10}$ E             &       244.8   &       2.11    &       -58.96  (0.10)  &       2.85    (0.22)  &       0.13    (0.01)  &       0.41    (0.03)  &       6.15    (0.07)  & D \\
271228.953      &       22$_{11,11}$-21$_{11,10}$ E             &       229.6   &       2.26    &       -58.91  (0.14)  &       3.31    (0.37)  &       0.16    (0.02)  &       0.57    (0.05)  &       8.00    (0.09)  & D \\
271253.478      &       22$_{11,12}$-21$_{11,11}$ E             &       229.6   &       2.27    &       -58.73  (0.06)  &       2.65    (0.22)  &       0.16    (0.02)  &       0.45    (0.03)  &       6.31    (0.07)  & D \\
271655.733      &       22$_{10,12}$-21$_{10,11}$ E             &       215,8   &       2.41    &       -58.52  (0.12)  &       3.56    (0.29)  &       0.19    (0.02)  &       0.71    (0.05)  &       9.40    (0.07)  & D \\
271680.635      &       22$_{10,13}$-21$_{10,12}$ E             &       215,8   &       2.41    &       -58.81  (0.10)  &       3.01    (0.24)  &       0.18    (0.02)  &       0.58    (0.04)  &       7.68    (0.07)  & D \\
273078.651      &       22$_{5,18}$-21$_{5,17}$ E               &       167.2   &       2.91    &       -58.73  (0.09)  &       3.39    (0.24)  &       0.21    (0.02)  &       0.77    (0.04)  &       8.52    (0.05)  & D \\
273095.053      &       22$_{5,18}$-21$_{5,17}$ A               &       167.2   &       2.91    &       -58.87  (0.09)  &       3.47    (0.23)  &       0.22    (0.02)  &       0.81    (0.05)  &       8.97    (0.06)  & D \\
274278.330      &       22$_{7,16}$-21$_{7,15}$ E               &       182.7   &       2.79    &       -58.74  (0.07)  &       3.39    (0.17)  &       0.21    (0.01)  &       0.75    (0.03)  &       8.73    (0.04)  & D \\
274285.334      &       22$_{7,16}$-21$_{7,15}$ A               &       182.7   &       2.81    &       -58.57  (0.07)  &       3.01    (0.19)  &       0.19    (0.01)  &       0.62    (0.03)  &       7.18    (0.05)  & D \\
\hline \hline
\end{tabular}
\begin{flushleft}
Notes: (1) Frequencies. (2) Transition. (3) Energy of the upper level. (4) Einstein
spontaneous emission coefficient. (5)-(8) Velocity, line width at half
intensity, brightness temperature, and integrated intensities. The one
sigma uncertainties are given in brackets. (9) Column densities of the
upper state level of the transition with respect to the upper state
 degeneracy $g_{\rm u}$. (10) D: detected lines.
\end{flushleft}
\end{table*}

\end{appendix}

\end{document}